\documentclass[11pt, a4paper]{article}
\usepackage{a4wide}
\usepackage{graphicx, afterpage}
\usepackage[numbers]{natbib}
\usepackage{enumitem}
\usepackage{hyperref}
\usepackage{setspace}
\usepackage{multirow}
\usepackage{amsmath, amsthm, amssymb, amsfonts}
\usepackage{bm}
\usepackage{color}     
\usepackage{pifont}
\usepackage{url}

\allowdisplaybreaks

\DeclareMathAlphabet\mathbfcal{OMS}{cmsy}{b}{n}

\newcommand{\mbf}{\mathbf}
\newcommand{\mc}{\mathcal}

\newcommand{\vep}{\varepsilon}

\renewcommand{\l}{\left}
\renewcommand{\r}{\right}

\def\wh{\widehat}
\def\wt{\widetilde}

\newcommand{\E}[0]{\mathsf{E}}
\newcommand{\Var}[0]{\mathsf{Var}}
\newcommand{\Cov}[0]{\mathsf{Cov}}

\newcommand{\R}{\mathbb{R}}

\newcommand{\N}{\mathbb{N}}
\newcommand{\iid}{\mbox{\scriptsize{iid}}}
\newcommand{\nn}{\nonumber}
\newcommand{\sic}{\mbox{SC}}
\newcommand{\rss}{\mbox{RSS}}

\newcommand{\bbI}{\mathbb{I}}
\newcommand{\cred}{\textcolor{black}}

\newcommand{\cp}{\theta}  
\newcommand{\Cp}{\Theta}  
\renewcommand{\c}{k} 
\newcommand{\C}{\mc K} 

\theoremstyle{definition}
\newtheorem{thm}{Theorem}[section]
\theoremstyle{definition}

\theoremstyle{definition}

\theoremstyle{definition}
\newtheorem{prop}[thm]{Proposition}
\theoremstyle{definition}

\theoremstyle{remark}

\theoremstyle{definition}
\newtheorem{defn}{Definition}[section]
\theoremstyle{definition}

\title{Data segmentation algorithms: Univariate mean change and beyond}
\author{Haeran Cho$^1$ and Claudia Kirch$^2$}

\begin{document}

\maketitle


\begin{abstract}
Data segmentation a.k.a.\ multiple change point analysis has received considerable attention due to its importance in time series analysis and signal processing, with applications in a variety of fields including natural and social sciences, medicine, engineering and finance.
 
The goal of this survey article is twofold: In the first part, we review the existing literature on the \emph{canonical data segmentation problem} which aims at detecting and localising multiple change points in the mean of univariate time series. We provide an overview of popular methodologies on their computational complexity and theoretical properties. In particular, our theoretical discussion focuses on the \emph{separation rate} relating to which change points are detectable by a given procedure, and the \emph{localisation rate} quantifying the precision of corresponding change point estimators, and we distinguish between whether a \emph{homogeneous} or \emph{multiscale} viewpoint has been adopted in their derivation.
We further highlight that the latter viewpoint provides the most general setting for investigating the optimality of data segmentation algorithms.

Arguably, the canonical segmentation problem has been the most popular framework to propose new data segmentation algorithms and study their efficiency in the last decades. In the second part of this survey, we motivate the importance of attaining an in-depth understanding of strengths and weaknesses of methodologies for the change point problem in a simpler, univariate setting, as a stepping stone for the development of methodologies for more complex problems. We illustrate this with a range of examples showcasing the connections between complex distributional changes and those in the mean.  We also discuss extensions towards high-dimensional change point problems where we demonstrate that the challenges arising from high dimensionality are orthogonal to those in dealing with multiple change points. 
\end{abstract}

\footnotetext[1]{School of Mathematics, University of Bristol, UK.
Email: \url{haeran.cho@bristol.ac.uk}. \\
Supported by the Leverhulme Trust Research Project Grant (RPG-2019-390).}

\footnotetext[2]{Department of Mathematics, Otto-von-Guericke University; Center for Behavioral Brain Sciences (CBBS); Magdeburg, Germany.
Email: \url{claudia.kirch@ovgu.de}.} 

\section{Introduction}

Change point analysis has a long tradition in statistics dating back to 1950s (see \citet{page1954}) 
and has been a very active field of research ever since due to its importance in time series analysis, 
signal processing and many other applications 
where data is routinely collected over time in naturally nonstationary environments.
\cred{This popularity can partly be explained by the fact that 
the assumption of piecewise stationarity underlying change point analysis,  
is one of the simplest forms of departure from  stationarity 
while at the same time, it is found to be reasonable for many applications. 
There are also other types of non-stationarities such as unit roots or local stationarity
that have been studied extensively in the literature which, however,
will not be discussed in this article.}


Research in change point analysis has predominantly focused on the following two directions: 
One line of research is on the development of sequential or online methodologies based on optimal stopping (see e.g.\ the recent books by \citet{tartakovsky2014sequential} and \citet{tartakovsky2019sequential}) 
which is being actively studied to date but will not be part of this review.
The second direction addresses the problem of change point detection in {\it a posteriori} or offline settings.
Much of the research in the last century dealt with change point testing 
against the at-most-one-change (AMOC) alternative
(see e.g.\ the books by \citet{csorgo1997} and \citet{chen2011parametric}).
Based on such tests, the location of a single change point can be estimated
with the optimal localisation properties,
typically by scanning for the point where the corresponding statistic attains its maximum.
More recent papers tackle change point problems in complex situations 
(see the review papers by \citet{aue2013} and \citet{horvath2014} 
and also Section~\ref{sec:ext} below), 
both sequentially and retrospectively,
and this review paper focuses on the latter. 

In retrospective change point analysis, 
it is often more realistic to allow for multiple change points, 
where the goal is to estimate both the total number and locations of the change points,
a.k.a.\ data segmentation.
\cred{While there already exist early papers showing the consistency of
various change point procedures that make use of e.g.\
binary segmentation
(i.e.\ iteratively testing for a change point and partitioning the data into two, \citep{vostrikova1981}),
or information criteria \citep{lee1995, yao1988},
the problem of detecting multiple change points in data sequences
experienced renewed interest in the statistical literature during the last 10 years.}
This was driven by several factors
including the emergence of scientific problems requiring high-precision and high-speed 
segmentation algorithms: Such examples can be found in
genomics \citep{chan2017, li2016, niu2012, olshen2004},
astronomy \citep{fisch2018}, neurophysiology \citep{messer2014}, climatology \citep{reeves2007},
cyber-security \citep{adams2016} and finance \citep{cho2012}, to name but a few.
Accordingly, there has been a surge of interest for 
computationally fast and statistically efficient methods for data segmentation,
and most effort in this direction has been devoted to
the problem of detecting multiple change points in the mean of univariate data.
In this article, we also take this as an important starting point
for developing change point methodologies in more complex situations.
We first review the methods for detection and localisation of 
multiple change points in the mean of univariate data,
which we refer to as the \emph{canonical segmentation problem},
complementing the articles reviewing and comparing different change point detection algorithms
such as \cite{aminikhanghahi2017, eckley2011, fearnhead2020, truong2020}.

As noted by \citet{brodsky2000} (see also the discussion in Section~\ref{section_beyondmean} below),
more complex change point problems
that allow changes in stochastic properties other than the mean,
can be reduced to the (possibly multivariate) canonical segmentation problem 
after applying a suitable transformation to the input data
that reveals the changes as those in the mean of the transformed data. 
Also, the challenges due to high dimensionality of the data, 
which is another important area of active research (see Section~\ref{section_HD} below),
are {\it orthogonal} to those arising from the presence of multiple change points over time. 
Thus we argue that the suite of algorithms developed for the canonical segmentation problem 
serve as a foundation for developing methodologies for more complex change point problems.

The rest of the paper is organised as follows. 
Section~\ref{sec:cp} describes the canonical segmentation problem
including its aim and theoretical benchmarks,
and Section~\ref{sec:seg} reviews the state-of-the-art methodologies developed for this purpose.
Section~\ref{sec:ext} provides an overview of the existing literature
on more complex change point problems beyond mean changes and finite dimensionality,
and discuss how the data segmentation algorithms described in Section~\ref{sec:seg}
can be adapted for such problems. 

\section{Multiple change point detection in the mean of univariate data}
\label{sec:cp}

We regard the following change point model
\begin{align}
X_t &= f_t + \vep_t = f_0 + \sum_{j = 1}^{q_n} d_{j} \cdot \bbI_{t \ge \cp_j + 1} + \vep_t, 
\quad \E(\vep_t) = 0 \text{ and } \Var(\vep_t) = \sigma^2,
\quad t = 1, \ldots, n, \label{eq:model} 
\end{align}
as the canonical change point model,
where we denote by $q_n$ the total number of change points in the mean of $\{X_t\}_{t = 1}^n$,
by $\cp_j = \cp_{j, n}, \, j = 1, \ldots, q_n$, their locations,
by $d_j = d_{j, n}$ the (signed) magnitude of change 
and by $\delta_j = \delta_{j, n} = \min(\cp_j - \cp_{j - 1}, \cp_{j + 1} - \cp_j)$ 
the minimum distance from $\cp_j$ to its neighbouring change points,
with the notational convention that $\cp_0 = 0$ and $\cp_{q_n + 1} = n$. 
In data segmentation, the goal is to consistently estimate
the number and the locations of the change points,
and we review the methodologies proposed for this purpose in Section~\ref{sec:seg}.

\cred{In contrast, 
in the corresponding testing problems, the aim is to test $H_0: \, q_n = 0$ against $H_1: \, q_n \ge 1$;
in the AMOC setting, the alternative hypothesis is given by $H_1: \, q_n = 1$.
There exist a wide variety of procedures proposed for change point testing,
not only against the alternative of mean shifts but also allowing for more complicated structural breaks
(see Section~\ref{section_beyondmean} for such examples). 
There are also a few methods that sequentially test for an increasing number of change points,
i.e.\ $H_0: \, q_n = m$ against $H_1: \, q_n = m + 1$ for given $m$
\citep{bai1998estimating, maruvsiakova2009tests}.
These tests can be used as a pre-processing step to determine whether the data is sufficiently close to stationary,
or whether a multiple change point detection methodology is required to segment the data
into (approximately) stationary stretches.
They can also serve as a basis for developing a data segmentation methodology,
with the simplest example being the binary segmentation algorithm, see Section~\ref{sec:loc} below.}

We assume that $\max_{1 \le j \le q_n} \vert d_j \vert = O(1)$
as well as $\min_{1 \le j \le q_n} \delta_j \to \infty$,
separating the change point detection problem under~\eqref{eq:model} 
from that of outlier detection.
Indeed, one of the main differences between the two problems is that
in outlier detection, an increase in sample size does not benefit
the performance of an outlier detection procedure beyond that 
it helps getting more precise estimators for the quantiles and,
even so, the size of the outlier is required to be larger than these quantiles for its detection.
On the other hand, in change point detection, 
the increased sample size (which corresponds to $\delta_j \to \infty$ under~\eqref{eq:model})
allows for the detection of local changes where $d_j \to 0$;
therefore, looking for outliers in the differenced time series $\{X_t - X_{t - 1}\}_{t = 2}^n$
cannot provide an adequate solution to the change point problem. 

Generally, the detectability of change points
is determined by an interplay between the magnitude of changes
and their distance to neighbouring change points:
Large changes are detectable even if they are close to the adjacent change points
while for smaller changes, they should be sufficiently distanced 
away from the neighbouring change points to be detectable.
When large frequent changes and small changes over long stretches 
are simultaneously present in the same time series,
it calls for a multiscale methodology that adapts to such situations and detects all change points. 
This gives rise to the distinction between a {\it homogeneous} change point scenario
where the magnitude of changes and the spacing between change points 
are on the same scale for all change points,
and the {\it multiscale} scenario where this is not the case.
The difficulty of the canonical segmentation problem 
for a given time series under~\eqref{eq:model},
is gauged by the detection lower bound $\Delta_n$ 
relating the magnitude of the change $d_j$ 
to the minimum distance to neighbouring change points $\delta_j$,
and different formulations of the detection lower bound 
reflect these different scenarios, 
see Definition~\ref{def_scenarios}~\ref{def:dlb} for the precise statement. 
When $\Delta_n$ diverges faster than the \emph{separation rate} associated with a given methodology,
changes are detectable with asymptotic power one and 
their locations can be estimated with accuracy. 

Frequently, a methodology is deemed consistent 
in multiple change point detection with the \emph{localisation rate} $\rho_n$,
when it returns
$\wh\Cp = \{\wh\cp_j, \, 1 \le j \le \wh q: \, \wh\cp_1 < \ldots < \wh\cp_{\wh q}\}$ satisfying
\begin{align}
\label{eq:consistency}
\wh q = q_n  \quad \text{and} \quad 
\max_{1 \le j \le q_n} w_j \vert \wh\cp_j \cdot \mathbb{I}_{j \le q_n} - \cp_j \vert \le \rho_n
\end{align}
with probability tending to one, for some $\rho_n/n \to 0$. 
The quantity $w_j$ denotes the difficulty associated with localising each change point $\cp_j$, 
which can also be formulated within the homogeneous or multiscale scenarios.

Definition~\ref{def_scenarios}, taken from \citet{cho2020}, distinguishes multiscale formulations
of detection lower bound and localisation rate from their non-multiscale, homogeneous counterparts. 
\begin{defn}
\label{def_scenarios} 
\begin{enumerate}[label = (\alph*)] 
\item \label{def:dlb} Detection lower bound:
\begin{enumerate}[label = (\roman*)]
\item \label{def:dlb:hom} \textbf{Homogeneous change points:} 
$\Delta_n = \min_{1 \le j \le q_n} d_j^2 \cdot \min_{1\le j \le q_n} \delta_j$.

\item \textbf{Finite mixture of homogeneous change points:} 
There are $N < \infty$ disjoint subsets of change points
with their indices given by $\mc J_k, \, k = 1, \ldots, N$, such that
$\bigcup_{k = 1}^N \mc J_k = \{1, \ldots, q_n\}$,
whereby change points within each subset are homogeneous as defined in~\ref{def:dlb:hom} 
and $\Delta_n = \min_{1 \le k \le N} (\min_{j \in \mc J_k} d_j^2 \cdot \min_{j \in \mc J_k}\delta_j$).
A special case is when there are finitely many changes with $N = q_n$.

\item \textbf{Multiscale change points:} $\Delta_n = \min_{1 \le j \le q_n} d_j^2 \, \delta_j$.
\end{enumerate}

\item \label{def:loc:rate} Localisation rate:
We distinguish between a \textbf{homogeneous localisation rate} 
where the estimation error for the $j$-th change point, $\vert \wh\cp_j - \cp_j \vert$,
is weighted globally with $w_j = \min_{1 \le j \le q_n} d_j^2$, and a \textbf{multiscale localisation rate} 
where it is weighted locally with $w_j = d_j^2$.
\end{enumerate}
\end{defn}

In the AMOC situation where $q_n \le 1$,
all scenarios in Definition~\ref{def_scenarios}~\ref{def:dlb} coincide
and reduce to the assumption \cred{$d_1^2 \min(\cp_1, n - \cp_1) \to \infty$} 
which is commonly found in the change point testing literature,
see Section~1.5 of \citet{csorgo1997}.
However, when $q_n > 1$ (and possibly diverging with $n$), they truly reflect different scenarios:
Proceeding from (i) to~(iii) 
enlarges the associated parameter space,
and thus (iii) provides the most general setting in which to
investigate the detection power of a segmentation procedure.
While most of the literature deals with the homogeneous setting (see Section~\ref{sec:overview}),
it is often too restrictive and does not permit large frequent jumps 
alternate with small jumps over long segments -- a situation naturally embedded in the multiscale setting.
In Definition~\ref{def_scenarios}~\ref{def:loc:rate},
the multiscale localisation rate relates the difficulty in accurate localisation of each change point
to the size of the corresponding jump. 

Propositions~\ref{prop:lb} and~\ref{prop:loc} state the benchmarks 
for the minimax optimal separation and localisation rates,
which have been established under the special case 
where $\{\vep_t\}_{t = 1}^n$ are i.i.d.\ (sub-)Gaussian random variables.
To the best of our knowledge, there do not exist equivalent results 
on the detection lower bound or the localisation rate (when $q_n \to \infty$) 
beyond the \cred{case of} i.i.d. sub-Gaussianity.
For the AMOC alternative,
it is well known that the minimax optimal localisation rate is given by $O_P(1)$ 
(see \citet{korostelev1987}), 
i.e.\ $\rho_n \to \infty$ arbitrarily slowly in~\eqref{eq:consistency}. 
This rate extends beyond Gaussianity and independence
when there are a finite number of change points,
see Corollary~4.4 of \citet{cho2020}.

\begin{prop}[{\bf Proposition~1 of \citet{arias2011}, lower bound on the minimax separation rate}] 
\label{prop:lb}
Under~\eqref{eq:model}, let $H_{0, n}: \, q_n = 0$ 
and $H_{1, n}$ describe the setting where $q_n= 2$, $d_n: = d_1 = -d_2$
and $\delta_n: = \cp_2 - \cp_1$ with $n^{-1} \delta_n \to 0$.
Then, $H_{0, n}$ and $H_{1, n}$ are asymptotically inseparable if
$\vert d_n \vert \sqrt{\delta_n} \le \sqrt{2\log(n/\delta_n)} - \nu_n$
where $\nu_n \to \infty$. 
\end{prop}

\begin{prop}[{\bf Proposition~6 of \citet{fromont2020}, 
lower bound on the minimax localisation rate for possibly an unbounded number of change points}]
\label{prop:loc}
Under~\eqref{eq:model} with $q_n \ge 2$, let $\vert d_j \vert =: d_n$ for all $j = 1, \ldots, q_n$,
and denote by 
\begin{align*}
\Xi = \l\{ (\cp_1, \ldots, \cp_{q_n}): \,
0 \equiv \cp_0 < \cp_1 < \ldots < \cp_{q_n} < \cp_{q_n + 1} \equiv n, \quad \text{and} \r.
\\*
\l. d_n^2 \min_{1 \le j \le q_n} (\cp_{j + 1} - \cp_{j-1}) > c \log(q_n) \r\}
\end{align*}
for some $c > 0$,
the parameter space for the locations of change points.
Then, for some $C > 0$, 
\begin{align}
\inf_{\C \in \N^{q_n}} \sup_{\Cp \in \Xi}
\E_{\Cp}\{d_H(\C, \Cp)\} \ge C d_n^{-2} \log(q_n) \nn
\end{align}
where $d_H(\C, \Cp) = \max\{\max_{\c \in \C} \min_{\cp \in \Cp} |\c - \cp|, 
\max_{\cp \in \Cp} \min_{\c \in \C} |\cp - \c|\}$
denotes the Hausdorff distance.
\end{prop}

\section{Review of data segmentation methodologies}
\label{sec:seg}

Broadly, algorithms  for the canonical segmentation problem can be categorised into two:
One line of research casts the problem as that of  {\it globally} optimising an objective function,
which imposes a penalty on the model complexity given e.g.\ by the number of change points;
the other is closely related to the change point testing literature,
whereby the test for detecting a single change point is applied {\it locally} 
for detection and estimation of multiple change points.
We briefly mention that there exist methods based on hidden Markov models 
where the main focus lies in modelling such phenomena, 
with algorithms for estimating the sequence of hidden states \citep{fuh2004, titsias2016}.
There are also methods allowing for Bayesian inference under change point models
\citep{ barry1993,du2016, fearnhead2006, rigaill2012,yao1984},
see \citet{eckley2011} for further references. 

In what follows, we review the literature on 
the canonical segmentation problem under~\eqref{eq:model}
according to the above categorisation,
providing brief descriptions of representative methodologies and their computational complexity
as well as theoretical properties 
in relation to the benchmark provided in Section~\ref{sec:cp}.

\subsection{Global optimisation methods}
\label{sec:global}

\subsubsection{$\ell_0$-penalisation}
\label{sec:ell:zero}

Information criteria have popularly been adopted in change point problems.
Under~\eqref{eq:model}, the Schwarz criterion \citep[SC]{schwarz1978} takes the form
\begin{align}
& \sic(\mc A) = \frac{n}{2} \log\l(\frac{\rss(\mc A)}{n}\r) + p_n(\mc A)
\quad \text{where} 
\label{eq:sc}
\\
& \rss(\mc A) = \sum_{j = 0}^m \sum_{t = k_j + 1}^{k_{j + 1}} (X_t - \bar{X}_{k_j:k_{j + 1}})^2
\quad \text{with} \quad
\bar{X}_{a:b} = \frac{1}{b - a} \sum_{t = a + 1}^b X_t.
\label{eq:rss}
\end{align}
The function $p_n(\mc A)$ denotes the penalty imposed on the complexity of a model
determined by a set of change point candidates 
$\mc A = \{k_j, \, 1 \le j \le m: \, k_1 < \ldots < k_m\} \subset \mc I_1 := \{1, \ldots, n - 1\}$
(with $k_0 = 0$ and $k_{m + 1} = n$);
a popular choice is $p_n(\mc A) = \lambda_n \; \vert \mc A \vert$
where $\lambda_n$ is a tuning parameter.
For some integer $N \le n$, 
we denote a collection of all subsets of $\mc I_1$ 
with their cardinality bounded by $N$, by
$\mathfrak{A}[N] = \l\{\mc A \subset \mc I_1: \, \vert \mc A \vert \le N \r\}$.
Then, minimising $\sic(\mc A)$ over $\mc A \in \mathfrak{A}[N]$ 
returns a set of change point estimators,
$\wh{\Cp}(\mathfrak{A}[N]) = \min_{\mc A \in \mathfrak{A}(N)} \sic(\mc A)$.

With the penalty linear in the number of candidate change points as
$p_n(\mc A) = (1 + \epsilon)\log(n) \cdot \vert \mc A \vert$,
the consistency of $\sic$ in estimating the total number of change points has been established
under i.i.d.\ Gaussianity, in \citet{yao1988} (with $\epsilon = 0$) 
and \citet{lee1995} (with a small constant $\epsilon > 0$),
in the setting where both $q_n$ and $N$ are fixed with $N > q_n$.
This result is complemented by \citet{yao1989}, where 
the consistency of the least squares estimators of $\cp_j,$ $1 \le j \le q_n$,
is established and their distribution is derived when $q_n$ is fixed and known.
Penalties depending on the spacing between the change points as well as the total number $\vert \mc A \vert$,
are considered in \citet{chen2006}, \citet{pan2006} and \citet{zhang2007},
and the application of the Akaike information criterion to the change point problem
is studied in \citet{ninomiya2005}.

\citet{boysen2009} consider the minimisation of the Potts functional
\begin{align}
\label{eq:potts}
\wh\Cp(\mathfrak{A}) = \arg\min_{\mc A \in \mathfrak{A}} P_0(\mc A),
\quad \text{where} \quad 
P_0(\mc A) = \frac{1}{n}\rss(\mc A) + p_n(\mc A)
\end{align}
with $p_n(\mc A) = \lambda_n \vert \mc A \vert$,
over $\mc A \in \mathfrak{A}[n - 1] =: \mathfrak{A}$, i.e.\
without any bound on the cardinality of $\mc A$ other than the trivial one.
Under i.i.d.\ sub-Gaussianity, 
they show the consistency of $\wh\Cp(\mathfrak{A})$
both in estimating the total number and the locations of change points,
provided that $\lambda_n \to 0$ and $n\lambda_n/\log(n) \to \infty$.
\citet{lavielle2000} establish the asymptotic consistency of 
the minimiser of the Potts functional under general conditions on $\{\vep_t\}$,
given that a suitable upper bound on the number of change points $N$ is available.
\citet{wang2018d} extend the results of \citet{boysen2009}
to the settings where $q_n \to \infty$, $\min_{1 \le j \le q_n} \vert d_j \vert \to 0$ and 
$n^{-1} \min_{1 \le j \le q_n} \delta_j \to 0$ as $n \to \infty$,
with the penalty $p_n(\mc A) = \log^{1 + \epsilon}(n) \cdot \vert \mc A \vert$ for some small $\epsilon > 0$.
By considering a location-dependent penalty
$p_n(\mc A) =  n^{-1} L\{\lambda \vert \mc A \vert + 2 \sum_{j = 0}^{m} \log(n/(k_{j + 1} - k_j))\}$
with some tuning parameters $L, \lambda > 0$,
\citet{fromont2020} show that the minimisation of $P_0(\mc A)$
achieves minimax optimality in both detection lower bound and change point localisation,
matching the rates provided in Propositions~\ref{prop:lb}--\ref{prop:loc}.

There exist dynamic programming algorithms 
which solve the $\ell_0$-penalised least squares estimation problem in~\eqref{eq:potts}.
The segment neighbourhood approach \citep{auger1989} 
requires an upper bound on the number of change points, say $N$,
and solves the constrained minimisation problems for a varying number of change points,
attaining the computational complexity of $O(N n^2)$.
The optimal partitioning \citep{jackson2005} approach, 
by recursively searching for the optimal partition 
conditioning on the latest change point estimator, 
achieves the computational complexity of $O(n^2)$ for a given choice of the penalty.
There are several methods which aim at speeding up the dynamic programming algorithms
by pruning the search space, such as 
the pruned exact linear time method \citep{killick2012}, 
the functional pruning technique \citep{rigaill2010}
and their extensions \citep{maidstone2017}.
\citet{haynes2017a} provide a methodology for finding the solution path 
for a range of penalty parameters $\lambda \in [\lambda_{\min}, \lambda_{\max}]$, 
when the penalty increases linearly in the size of $\mc A$
as $p_n(\mc A) = \lambda \vert \mc A \vert$,
while \citet{lebarbier2005detecting} proposes the slope heuristic method
that examines how the segmentation result varies with such $\lambda$.
\citet{tickle2020} investigate the parallelisation of dynamic programming-based methods
for fast computation when $n$ is very large.
\cred{Imposing an autoregressive (AR) structure of order one on $\{\vep_t\}$,
\citet{chakar2017robust} propose an adaptation of~\eqref{eq:sc}
and provide an algorithm for approximating its global minimiser.
Further allowing for random walk-like fluctuations, 
\citet{romano2021detecting} provides an  efficient dynamic programming algorithm
that returns an exact solution of the minimisation problem.}

Finally, we mention that the objective function 
based on the minimum description length principle \citep{grunwald2004,rissanen1978}
imposes an $\ell_0$-type penalty on the model complexity,
and has successfully been applied to piecewise parametric modelling
\citep{aue2014, davis2013, davis2006}
in combination with the genetic algorithm to traverse the parameter space
and solve the optimisation problem.

\subsubsection{$\ell_1$-penalisation} 
\label{sec:ell:one}
 
While efficient algorithms exist for the optimisation of $\ell_0$-penalised objective functions,
the computational complexity typically scales quadratically in $n$.
Relaxing the $\ell_0$-penalty in~\eqref{eq:potts} 
to an $\ell_1$-penalty (a.k.a.\ total variation penalty)
gives rise to 
\begin{align}
\label{eq:fused}
P_1(\mbf g) &= \frac{1}{n} \sum_{t = 1}^n (X_t - g_t)^2 + \lambda_n \sum_{t = 1}^{n - 1} \vert g_{t + 1} - g_t \vert,
\quad \mbf g = (g_1, \ldots, g_n)^\top, 
\end{align}
which is closely related to the fused Lasso,
originally introduced in \citet{tibshirani2005}
to take into account the ordering of the features in the data.
The locations of change points are estimated by the locations of jumps in 
$\wh{\mbf g} = \arg\min_{\mbf g \in \R^n} P_1(\mbf g)$.
\citet{harchaoui2010} show that adopting the least angle regression 
proposed by \citet{efron2004} in the context of $\ell_1$-penalised linear regression,
the computational complexity of solving for $\wh{\mbf g}$ is $O(N^3 + N n \log(n))$,
where $N$ denotes the maximal number of change point candidates to be considered and needs to be specified.
The properties of the $\ell_1$-penalised least squares estimator $\wh{\mbf g}$,
both in its $\ell_2$-error in estimating the mean and change point localisation,
has been studied in \citet{rinaldo2009} and \citet{harchaoui2010} under i.i.d.\ (sub-)Gaussianity.
\citet{lin2017} provide a near-minimax optimal rate (up to $\log\log(n)$) 
for the $\ell_2$-error when the change points are sparse,
but also show that $\wh{\mbf g}$ tends to over-estimate the number of change points
and requires further post-processing for consistent estimation of~$q_n$.

\subsubsection{Multiscale change point segmentation methods}
\label{sec:mcps}

\citet{li2019} define a class of
multiscale change point segmentation (MCPS) methods,
which include the simultaneous multiscale change point estimators
controlling for the family-wise error rate \citep[SMUCE]{frick2014}
and the false discovery rate \citep[FDRseg]{li2016} in estimating the number of change points.
They solve a (non-convex) optimisation problem,
which searches for the most parsimonious candidate estimator of the piecewise constant function
over the acceptance region of a suitable multiscale test. 
More specifically, the optimisation problem is given by
\begin{align}
\label{eq:mcps}
\min_{g \in S[0, 1)} \# J(g) \quad \text{subject to} \quad
\max_{\substack{I \in \mc M \\ g \equiv \gamma_I \text{ on } I}} 
\l\{ \frac{1}{\sqrt{\vert I \vert}} \l\vert \sum_{t/n \in I} (X_t - \gamma_I) \r\vert - s_I \r\} \le \eta,
\end{align}
where $\# J(g)$ denotes the number of jumps in $g \in S[0, 1)$,
the space of right-continuous step functions on $[0, 1)$
(whose change points lie on the grid $\{i/n\}_{i = 0}^{n - 1}$),
$\mc M$ a multiscale system of sub-intervals of $[0, 1)$ fulfilling a set of requirements,
$s_I$ a scale penalty 
and $\eta$ a threshold controlling the over-estimation probability. 

\citet{frick2014} establish the near-minimax optimality of  SMUCE
as given in Propositions~\ref{prop:lb}--\ref{prop:loc},
under the exponential family regression model with the parameter
varying over time in a piecewise constant manner,
and discuss an efficient implementation of the optimisation problem in~\eqref{eq:mcps} 
via dynamic programming.
For the canonical segmentation problem,
\citet{pein2015} and \citet{dette2018} address the issues arising from
heteroscedasticity and serial dependence in $\{\vep_t\}_{t = 1}^n$, respectively.

\subsubsection{\cred{Model selection-based segmentation}}
\label{sec:cv}

\cred{\citet{arlot2011segmentation} investigate the use of cross-validation 
for the change point problem under~\eqref{eq:model}.
With $\rss(\mc A)$ defined in~\eqref{eq:rss} serving 
as the empirical risk measure of a change point model $\mc A$,
the proposed method adopts the leave-one-out or $K$-fold cross-validation methods 
for final model selection, in place of imposing a penalty on the model complexity as
the class of methods discussed in Sections~\ref{sec:ell:zero} and~\ref{sec:ell:one}.
The paper provides a dynamic programming approach to efficiently compute the procedure
and studies its theoretical performance in terms of the $\ell_2$-error.}

\subsection{Local testing methods}
\label{sec:loc}

The methodologies belonging to this class generally proceed in two stages: 
First, the data is scanned for candidate change point estimators,
from which the final estimators are selected via a model selection method.


In the first step, candidate generation is achieved by 
multiscale scanning of the data for change points 
using the following weighted cumulative sum (CUSUM) statistics
computed on the segment $\{X_t, \, s < t \le e\}$:
\begin{align*}
\mc T_{s, k, e}(X) = \sqrt{\frac{(k - s)(e - k)}{e - s}}\l(\bar{X}_{s:k} - \bar{X}_{k:e}\r) \quad \text{where} \quad
\bar{X}_{a:b} = \frac{1}{b - a} \sum_{t = a + 1}^b X_t,
\end{align*}
for $0 \le s < k < e \le n$.
In the AMOC setting, 
it is well-known that the location of the maximum of the CUSUM statistics given by
$k_{\circ} = \arg\max_{0 < k < n} \vert \mc T_{0, k, n}(X) \vert$,
is a consistent minimax optimal estimator for the change point.
This motivates taking the maximiser of the CUSUM statistics 
as a change point candidate within the local interval $(s, e)$
provided that $\max_{s < k < e} \vert \mc T_{s, k, e}(X) \vert$ is large enough. 

A variety of methods adopt $\mc T_{s, k, e}(X)$ 
in order to scan the data for change points
with a method-specific search space 
$\mc S \subset \mc I_3 := \{(s, k, e): \, 0 \le s < k \le n\}$,
which is either fixed, random or data-dependent.
The construction of the search space aims at isolating
change points for their detection and localisation:
Ideally, for each change point, 
$\mc S$ contains at least one interval which is as large as possible
for the detection of even a small change,
while containing only a single change point for its localisation
and thus avoiding the contamination by multiple change points 
(further discussion on this point is found in Section~\ref{sec:bs}).
Once candidate change point estimators are identified, 
it remains to determine the final set of estimators
which is achieved by a model selection (pruning) methodology,
such as thresholding or information criteria.

In Sections~\ref{sec:bs}--\ref{sec:bu}, we discuss
candidate generation strategies which scan the data
using $\mc T_{s, k, e}(X)$ with different search spaces $(s, k, e) \in \mc S$.
Then in Section~\ref{sec:ms}, we describe popular model selection methodologies
applied to the thus-generated candidate change point estimators.

\subsubsection{Binary segmentation algorithm and its extensions}
\label{sec:bs}

Binary segmentation (BS) algorithm \citep{scott1974, vostrikova1981} 
is a generic method for multiple change point estimation
which recursively partitions the data into two.
For multiple change point detection under~\eqref{eq:model}, 
it identifies a candidate estimator $k_1$
from scanning the CUSUM statistic
$\vert \mc T_{0, k, n}(X) \vert, \, 1 \le k < n$, as its maximiser.
If the maximum CUSUM (max-CUSUM) statistic $\mc T_{0, n}^{\max}
= \max_{1 \le k < n} \vert \mc T_{0, k, n}(X) \vert$ is sufficiently large,
the data is partitioned into $\{X_t\}_{t = 1}^{k_1}$ and $\{X_t\}_{t = k_1 + 1}^n$,
and the scanning and maximisation of the CUSUM statistics 
is repeated over each partition separately until a stopping criterion is met.
The search space of BS is stochastic, as the search space at the $i$-th iteration with $i \ge 2$
depends on the change points detected at earlier iterations,
and its complexity is typically $O(n\log(n))$.
\citet{venkatraman1992} established the consistency of BS
while allowing the number of change points, $q_n$, to diverge with $n$.

\begin{figure}
\centering
\includegraphics[width = .9\linewidth]{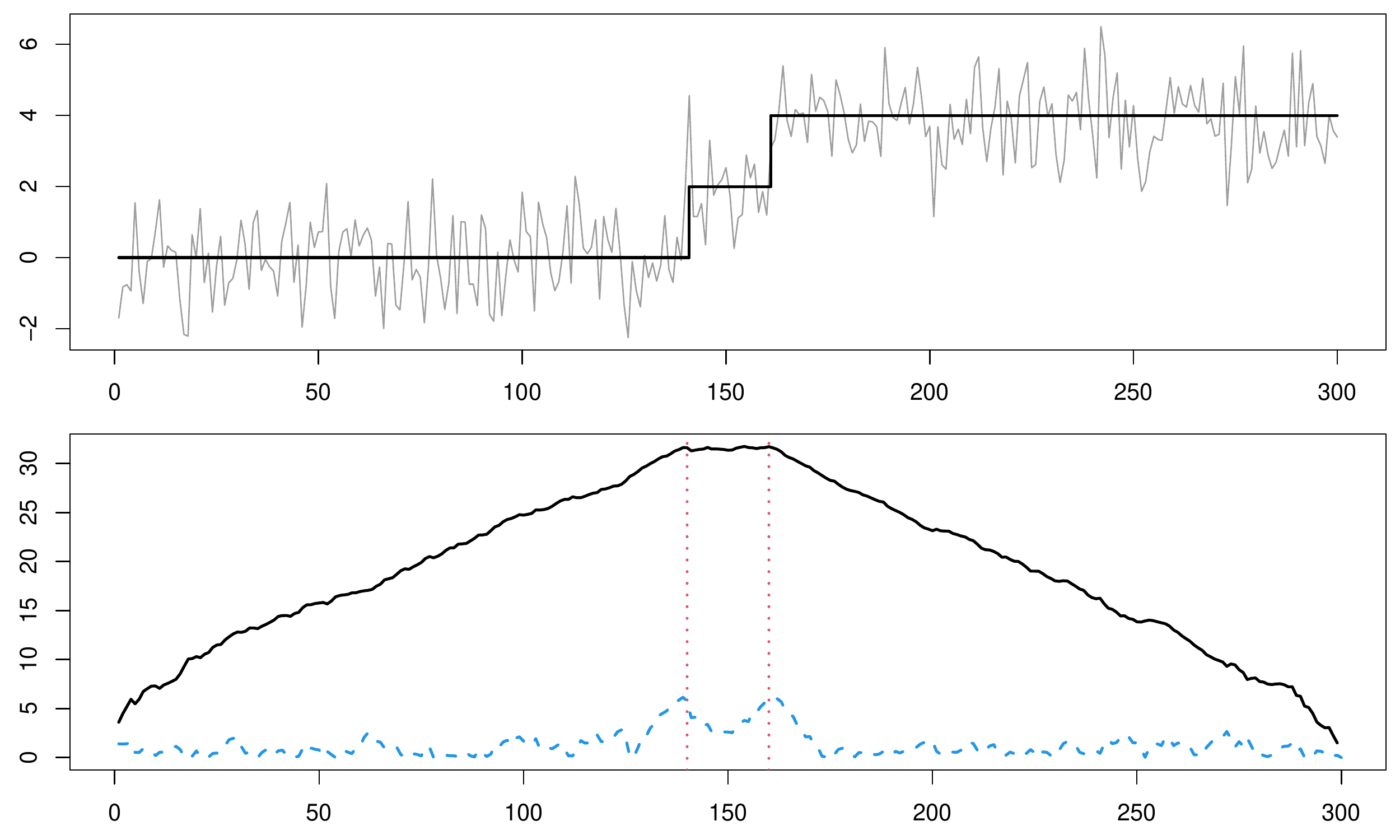}
\caption{Top: A realisation $\{X_t\}_{t = 1}^n$ from~\eqref{eq:model}
with $f_t = 2 \cdot \mathbb{I}_{t \ge 141} + 2 \cdot \mathbb{I}_{t \ge 161}$,
$\vep_t \sim_{\iid} \mc N(0, 1)$ and $n = 300$
and the piecewise constant signal $f_t$ (bold).
Bottom: $\vert \mc T_{0, k, n}(X) \vert, \, 1 \le k < n$ (solid),
and $\vert \mc T_{k - G, k, k + G}(X) \vert, \, k \in \mc S_{\text{M}}(G)$
with a boundary correction and $G = 15$ (broken, see Section~\ref{sec:mosum} for further details),
plotted with the vertical dotted lines denoting the locations of $\cp_j, \, j = 1, 2$.}
\label{fig:ex}
\end{figure}

The CUSUM statistic coincides with the likelihood ratio statistic for
testing the null hypothesis of no change point ($H_0: \, q_n = 0$) 
against the AMOC alternative ($H_1: \, q_n = 1$)
under i.i.d.\ Gaussian errors, and
thus is particularly well-adopted for the detection and localisation of a single change point.
On the other hand, in the presence of multiple change points, 
BS often performs poorly in estimating the locations of the change points
as the search space may include intervals 
contaminated by more than one change points.
The CUSUM-based test still attains good power 
when applied to such intervals,
but the statistic may be unsuitable for locating the change points due to this contamination.
Figure~\ref{fig:ex} provides an illustration where, while 
$\vert \mc T_{0, k, n}(X) \vert, \, 1 \le k < n$, attain large values near the two change points,
their precise locations are not well pointed out by the statistic.
In such a situation, procedures that aim at isolating 
each change point for its detection and localisation,
such as the moving sum procedure (see Section~\ref{sec:mosum}),
perform better in locating the change points
although they tend to lack power by comparison when used solely for testing.

The wild binary segmentation (WBS) algorithm proposed in \citet{fryzlewicz2014}
sets out to remedy the limitation of BS
by scanning the data over a random grid
$\mc R_n \subset \mc I_2 := \{(s, e): \, 0 \le s < e \le n\}$,
which amounts to examining 
$\mc T_{s, k, e}(X), \, (s, k, e) \in \mc S_{\text{W}}(\mc R_n)$ with
\begin{align*}
\mc S_{\text{W}}(\mc R_n) = \l\{ (s, k, e): \, k = s + 1, \ldots, e, \, (s, e) \in \mc R_n\r\}.
\end{align*}
As in BS, WBS recursively identifies candidate change point estimators 
and partitions the data into two, 
but the search is performed over the randomly drawn background intervals
and thus its computational complexity is determined by the choice of $\mc R_n$ as $O(n \vert \mc R_n \vert)$.
In order to ensure that each change point is considered on its own within a large enough interval
for its detection and estimation, WBS requires that a sufficiently large number of 
background intervals are drawn;
more precisely, $\vert \mc R_n \vert$ is required to increase as 
the minimum distance between the change points, $\delta_n := \min_{1 \le j \le q_n} \delta_j$,
decreases such that $(n/\delta_n)^{-2} \vert \mc R_n \vert \to \infty$.
\citet{fryzlewicz2014} refines the theoretical analysis of \citet{venkatraman1992},
and shows that WBS outperforms BS in terms of 
both the separation and the localisation rates.

There exist several modifications and extensions of WBS.
\citet{wang2018d} consider WBS with the lengths of the background intervals in $\mc R_n$
bounded from above by $\delta_n$,
which ensures that each background interval contains a single change point at most,
leading to an improvement of the localisation rate.
\citet{baranowski2019} propose a change point-like feature detection method termed 
the narrowest-over-threshold (NOT) where, at each recursive iteration,
the shortest interval $(s, e]$ 
over which the associated max-CUSUM statistic, $\max_{s < k < e} \vert \mc T_{s, k, e}(X) \vert$,
exceeds a given threshold is selected, rather than
the interval returning the largest max-CUSUM.
\citet{fryzlewicz2020} proposes WBS2, which iteratively draws random intervals
and produces a complete solution path,
i.e.\ an ordered sequence of $\mc I_1 = \{1, \ldots, n - 1\}$ 
according to their importance measured by the associated CUSUM statistics.
\citet{kovacs2020} investigate the seeded binary segmentation (SBS), 
which generates the search grid $\mc R_n$ in a deterministic fashion.
In \citet{lu2017}, a sampling strategy is investigated in combination with BS and WBS 
for handling very large $n$.

In all above, the theoretical investigation is focused on the case of uncorrelated data,
while \citet{cho2020b} discuss the usefulness of the WBS-like strategy
in separating the effects of change points and time series fluctuations
in the presence of serial correlations.
Finally, we mention the circular binary segmentation 
proposed in \citet{venkatraman2007} for the analysis of array CGH data,
where the test statistics are tailored for detecting changes of epidemic nature,
where $f_{\cp_{2j + 1}} = f_{\cp_{2j +2} + 1}$ for $j = 0, \ldots, \lfloor q_n/2 \rfloor$ under~\eqref{eq:model}.

\subsubsection{Moving sum procedures}
\label{sec:mosum}

The moving sum (MOSUM) procedure has been adopted
for simultaneously detecting and locating multiple change points
\citep{chu1995, eichinger2018, huvskova2001}.
For the canonical segmentation problem it scans the data for change points
using the statistics $\mc T_{s, k, e}(X)$ over a fixed grid $(s, k, e) \in \mc S_{\text{M}}(G)$,
which is fully determined by the choice of a bandwidth $G = G_n$ as
\begin{align}
\label{eq:mosum}
\mc S_{\text{M}}(G) = \l\{(s, k, e): \, s = k - G, \, e = k + G, \, k = G ,\ldots, n - G\r\}.
\end{align}
The search space is of cardinality $\vert S_{\text{M}}(G) \vert = O(n)$
since for every time point $k$, only the local interval $(s, e) = (k - G, k + G)$ is considered;
we sometimes denote that $k \in \mc S_{\text{M}}(G)$ where there is no confusion.
\citet{eichinger2018} propose two criteria for the simultaneous localisation of the change points
whereby each estimator is obtained as 
a local maximiser of $\vert \mc T_{s, k, e}(X) \vert, \, k \in \mc S_{\text{M}}(G)$,
over a sufficiently large interval relative to $G$, 
at which the MOSUM statistic exceeds a theoretically motivated threshold.

\citet{messer2014, messer2018}
consider a multiscale extension of the MOSUM procedure
with the search space
\begin{align*}
\mc S_{\text{M}}(\mc G) = \l\{ (s, k, e): \, s = k - G_h, \, e = k + G_h, \, k = G_h, \ldots, n - G_h; 
\, h = 1, \ldots, H\r\}
\end{align*}
where $\mc G = \{G_1, \ldots, G_H\}$ denotes a set of bandwidths.
The screening and ranking procedure of \citet{niu2012} 
also adopt a multiscale MOSUM procedure for candidate generation. 

The use of asymmetric bandwidths $\mbf G = (G_\ell, G_r)$
with $\mc S_{\text{M}}(\mbf G) =\{(s, k, e): \, 
s = k - G_\ell, \, e = k + G_r, \, k = G_\ell, \ldots, n - G_r\}$
has been investigated by \citet{meier2021mosum}.
\citet{cho2020} propose the following scheme for asymmetric bandwidth selection
based on the Fibonacci series $\{F_m\}$
constructed as $F_m = F_{m - 1} + F_{m - 2}, \, m \ge 2$ with $F_0 = F_1 = 1$:
The set of asymmetric bandwidths is determined as
\begin{align*}
\mc H_n = \l\{ (G_\ell, G_r): \, G_{\ell \slash r} = G_{\min} \cdot F_{\ell \slash r}, \, 1 \le \ell, r \le H_n, 
\quad \text{and} \quad
\frac{\max(G_\ell, G_r)}{\min(G_\ell, G_r)} \le C_{\text{asym}} \r\},
\end{align*}
with $H_n$ chosen such that $G_{H_n} < \lfloor n/\log(n)\rfloor$
for some minimum allowable bandwidth $G_{\min}$
and a fixed constant $C_{\text{asym}} \ge 1$.
The latter constant controls the unbalancedness of 
the asymmetric bandwidth $(G_\ell, G_r)$ such that
the computational complexity of the multiscale MOSUM procedure
involving $T_{s, k, e}(X), \, (s, k, e) \in \mc S_{\text{M}}(\mc H_n)$ is $O(n\log(n))$.

The multiscale application of the MOSUM procedure necessitates a pruning procedure
for the removal of conflicting estimators detected for the identical change points at multiple bandwidths,
as well as possibly spurious estimators detected at small bandwidths. 

\subsubsection{Bottom-up and sequential approaches}
\label{sec:bu}

There exist data segmentation methods that set out to isolate each change point
by considering the data from finer to coarser scales.
\citet{fryzlewicz2017} propose a method based on 
the tail-greedy unbalanced Haar (TGUH) transform,
which performs a bottom-up data decomposition using the unbalanced Haar wavelet basis.
Starting from the finest grid $\mc S^{(0)} = \{1, 2, \ldots, n - 1\}$,
it identifies the regions (defined by the adjacent elements of the grid) to be merged 
as where the corresponding statistics $\l\vert \mc T_{s, k, e}(X) \r\vert$
computed over $(s, k, e) = (k^{(0)}_{i - 1}, k^{(0)}_{i}, k^{(0)}_{i + 1}), \, k^{(0)}_i \in \mc S^{(0)}$
(with $k^{(\ell)}_0 = 0$ and $k^{(\ell)}_{|\mc S^{(\ell)}| + 1} = n$ for all $\ell \ge 0$) 
are small, and accordingly update the search grid. 
Repeatedly applying the above steps
until all $\vert \mc T_{s, k, e}(X) \vert$ are large enough
over $(s, k, e) = (k^{(\ell)}_{i - 1}, k^{(\ell)}_{i}, k^{(\ell)}_{i + 1}), \, k^{(\ell)}_i \in \mc S^{(\ell)}$
for some~$\ell$,
we obtain an ordering of change point candidates with an implicit tree-like structure
on which model selection is performed.
By greedily selecting the regions to be merged, which is managed by a parameter
controlling their proportion out of the remaining regions,
TGUH achieves a computational complexity of $O(n\log^2(n))$.
A similar idea of bottom-up change point analysis 
has also been explored in the reverse segmentation proposed by \citet{chan2017}, 
and the backward procedure of \citet{shin2020}.

The pseudo-sequential procedure of \citet{venkatraman1992}
adapts an online change point detection algorithm to the offline data segmentation problem.
The Isolate-Detect (IDetect) proposed in \citet{anastasiou2019} operates in a similar vein,
sequentially searching for change points from both ends.
More specifically, it scans $\mc T_{s, k, e}(X), \, (s, k, e) \in \mc S_L(\ell, r) \cup \mc S_R(\ell, r)$,
where $\mc S_L(\ell, r)$ and $\mc S_R(\ell, r)$ 
denote the search spaces containing the intervals expanded from the left to the right, 
and those expanded from the right to the left, respectively,
for given $0 \le \ell < r \le n$, as
\begin{align*}
\mc S_L(\ell, r) &= \l\{(s, k, e): \, \ell = s < k < e \le r \text{ with } e = m\lambda_n, \, m \in \mathbb{N} \r\}, 
\\
\mc S_R(\ell, r) &= \l\{(s, k, e): \, \ell \le s < k < e = r \text{ with } s = n - m\lambda_n, \, m \in \mathbb{N} \r\},
\end{align*}
where $\lambda_n$ denotes a parameter satisfying $\lambda_n \le \delta_n = \min_{1 \le j \le q_n} \delta_j$.
Starting with $(\ell, r) = (0, n)$, it iteratively updates $(\ell, r)$ with 
the previously detected change point estimators. 
The choice of $\lambda_n$ involves balancing between the computational complexity
$O(n^2\lambda_n^{-1})$, 
and the change point localisation property
(smaller $\lambda_n$ providing better guarantee of isolating each change point
for their detection and localisation).

\subsubsection{Model selection}
\label{sec:ms}

The methods outlined in Sections~\ref{sec:bs}--\ref{sec:bu}
produce a set of candidate change point estimators $\mc C$
from scanning the data at multiple scales;
some return a subset of $\mc I_1 = \{1, \ldots, n - 1\}$,
while the others generate a (complete) solution path ordering 
the elements of $\mc I_1$ according to their importance.
For the estimation of the total number and the locations of the change points,
it remains to select the final model from the set of candidate estimators. 

Closely related to the change point testing literature, 
thresholding is popularly adopted for model selection in change point detection.
\citet{eichinger2018} derive the (distributional) limit behaviour of the MOSUM test statistic 
$\max_{k \in \mc S_{\text{M}}(G)} \vert \mc T_{s, k, e}(X) \vert$ under the null hypothesis of no change,
under general conditions allowing for heavy tails and serial dependence in $\{\vep_t\}$
with the bandwidth $G$ satisfying $G \to \infty$ while $G/n \to 0$;
the result is extended for the asymmetric MOSUM statistic in \citet{meier2021mosum}.
\citet{messer2014, messer2018} approximate the MOSUM statistic 
by a Gaussian process when $G$ grows linearly with $n$.
With such asymptotic null distributions at hand,
we can derive a theoretically motivated threshold at a chosen significance level
and control for the family-wise error rate.

For methods such as BS, WBS and their extensions,
the test statistics associated with change point estimators are
dependent on the previously detected ones, and thus 
it is challenging to investigate their asymptotic null distributions.
Instead, the thresholds are often given by a range depending on 
$n$, the level of noise and other typically unknown quantities 
such as $\min_j \vert d_j \vert$ and $\min_j \delta_j$,
see \citep{anastasiou2019, baranowski2019, fryzlewicz2014, fryzlewicz2017, venkatraman1992, wang2018d}.
\citet{fang2020} provide a detailed analysis of the tail probability of 
$\max_{(s, k, e) \in \mc I_3} \vert \mc T_{s, k, e}(X) \vert$
under i.i.d Gaussianity, which enables the selection of the threshold at a prescribed significance level.

For a critical value or a threshold to be useful,
we need to estimate the level of noise given by
the error variance $\sigma^2 = \Var(\vep_t)$ under independence
or the long-run variance $\tau^2 = \Var(\vep_t) + 2 \sum_{h = 1}^\infty \Cov(\vep_0, \vep_h)$
when serial dependence is permitted.
Several estimators exist for both $\sigma^2$ and $\tau^2$ 
under the model~\eqref{eq:model} \citep{axt2020, dette2018, tecuapetla2017},
but the difficulty associated with the estimation of the latter
has been well-documented particularly in the presence of multiple change points \citep{perron2006},
and the former also becomes challenging to estimate when frequent change points are present.
As a remedy, \citet{fryzlewicz2020} proposes the
`steepest drop to low levels' (SDLL) model selection methodology
applicable with a complete solution path ordering $\mc I_1 = \{1, \ldots, n - 1\}$, 
with thresholding adopted as a secondary criterion.

Information criteria such as those discussed in Section~\ref{sec:ell:zero},
have frequently been adopted for final model selection
in combination with the methods belonging to the local testing category.
\citet{fryzlewicz2014} propose to minimise the strengthened SC,
defined in~\eqref{eq:sc} with the penalty function 
$p_n(\mc A) = \log^{1+\epsilon}(n) \cdot \vert \mc A \vert$ for some small $\epsilon > 0$,
along a solution path 
$\{k_1\} \subset \{k_1, k_2\} \subset \ldots \subset \{k_1, \ldots, k_N\} \subset \mc I_1$
returned by WBS with some pre-determined $N > q_n$;
the subset that minimises the criterion is chosen as the final model.
A similar approach is also taken by
\cite{anastasiou2019, baranowski2019, niu2012}.

Alternatively, \citet{cho2020} propose a top-down pruning methodology based on SC, 
which searches for the final model $\wh{\Theta}$ 
from a set of candidate change point estimators, say $\mc C$, according to the following rules:
(i) Adding further estimator to $\wh{\Theta}$ monotonically increases SC, and
(ii) among all subsets satisfying~(i), $\wh{\Theta}$ attains the minimal cardinality
and (if there exist ties) the minimum SC. 
This approach gains computationally
compared to minimising the SC exhaustively among all the $2^{\vert \mc C \vert}$ subsets of $\mc C$,
and achieves consistency in multiple change point estimation
under general assumptions permitting dependence and heavy tails in $\{\vep_t\}$.
A localised version of the pruning method attaining linear complexity is proposed 
for further computational consideration,
and it is shown to match the minimax optimal multiscale separation and localisation rates
when applied to the candidates generated by a multiscale MOSUM procedure.
\citet{cho2020b} explore the top-down approach 
under a parametric modelling assumption on $\{\vep_t\}$,
which effectively avoids the difficult task 
of evaluating the information criterion at a model under-specifying the number of change points.

\citet{messer2014, messer2018} consider a bottom-up merging method
for the multiscale MOSUM procedure:
Accepting all the estimators from the smallest bandwidth, 
it proceeds to coarser scales and only accepts a change point estimator 
if its detection interval does not contain any estimators that are already accepted. 
A similar strategy, named the local segmentation, has also been considered in \citet{chan2017},
which permits asymmetric bandwidths.

\subsection{An overview of change point detection methodologies}
\label{sec:overview}

\begin{table}[htb]
\caption{Comparison of change point detection methodologies on
their detection lower bound
and the separation and localisation rates derived under i.i.d.\ (sub-)Gaussianity,
where $d_n = \min_{1 \le j \le q_n} \vert d_j \vert$ and $\delta_n = \min_{1 \le j \le q_n} \delta_j$.
Also, we provide the computational complexity of different methods
with $N$ denoting the maximum allowable number of change points where relevant,
and list their software implementations.
The results reported here for BS and WBS are taken from \cite{cho2015c, wang2018}.}
\label{table:overview}
\resizebox{\columnwidth}{!}{
\begin{tabular}{ll | cc | cc | c | c}
\hline\hline
&& \multicolumn{2}{|c|}{Detection lower bound} & \multicolumn{2}{|c|}{Localisation} & Computational &  \\
&& $\Delta_n$ & Separation rate & $w_j$ & Rate & complexity & Implementations \\ \hline
$\ell_0$- & \cite{fromont2020} & $\min_j d_j^2 \delta_j$ & $\log(n/\delta_n)$ & $d_j^{2}$ & $\log(q_n)$  &  $O(n^2)$ & \citep{changepoint, s3ib, fpop} \\
penalisation & \cite{wang2018d} & $d_n^2\delta_n$ & $\log(n)$ & $d_j^{2}$ & $\log(n)$ & $O(n^2)$ \\
\hline
$\ell_1$- & \cite{harchaoui2010} & -- & -- & 1 & $\log^2(n)$  &  $O(N^3 + $ & -- \\
penalisation & \cite{lin2017} & -- & -- & $d_n^2$ & $\log(n)\log\log(n)$ & $Nn\log(n))$ & \\
\hline
MCPS & \cite{frick2014} & $d_n^2\delta_n$ & $\log(n/\delta_n)$ & $d_n^{2}$ & $\log(n)$ & $O(n^2)$ & 
\citep{stepR, fdrseg} \\
& \cite{li2019} & $d_n^2\delta_n$ & $q_n\log(n)$ & $d_n^{2}$ & $q_n\log(n)$ & -- \\
\hline
BS & \cite{fryzlewicz2014} & $d_n^2 \delta_n$ & $n^3\delta_n^{-5/2}\log(n)$ & 
$d_n^{2}$ & $(n/\delta_n)^5 \log(n)$ & $O(n\log(n))$ & \citep{changepoint} \\
\hline
WBS & \cite{fryzlewicz2014} & $d_n^2 \delta_n$ & $(n/\delta_n)^4\log(n)$  & 
$d_j^{2}$ & $(n/\delta_n)^4\log(n)$ & $O(n\vert \mc R_n \vert)$ &
\citep{breakfast} \\
& \cite{wang2018d} & $d_n^2 \delta_n$ & $\log(n)$ & $d_j^{2}$ & $\log(n)$ & with 
\\
\cline{1-6}
NOT & \cite{baranowski2019} & $d_n^2 \delta_n$ & $\log(n)$ & $d_j^{2}$ & $\log(n)$ &
$(n/\delta_n)^2/\vert \mc R_n \vert \to 0$ 
\\
\hline
MOSUM & \cite{eichinger2018} & $d_n^2\delta_n$ & $\log(n/\delta_n)$ & $d_j^{2}$ & $\log(q_n)$ & $O(n)$
& \citep{mosum} \\
 & \cite{cho2020} & $\min_j d_j^2 \delta_j$ & $\log(n)$ & $d_j^{2}$ & $\log(q_n)$ & $O(n\log(n))$ \\
 & \cite{chan2017} & $d_n^2\delta_n$ & $\log(n/\delta_n)$ & $d_j^{2}$ & $\log(n)$ & $O(n\log(n))$ \\
\hline
IDetect & \cite{anastasiou2019} & $d_n^2\delta_n$ & $\log(n)$ & $d_j^{2}$ & $\log(n)$ & $O(n^2/\delta_n)$
& \citep{breakfast, IDetect} \\
TGUH & \cite{fryzlewicz2017} & $d_n^2\delta_n$ & $\log^2(n)$ & $d_n^{2}$ & $\log^2(n)$ & $O(n\log^2(n))$ \\
\hline
\end{tabular}}
\end{table}

Table~\ref{table:overview} provides a comprehensive overview of 
change point methodologies discussed in Sections~\ref{sec:global}--\ref{sec:loc},
which expands upon Table~1 of \citet{cho2020}.
They are compared on their theoretical properties 
such as how the detection lower bound is formulated,
their separation and localisation rates under i.i.d.\ (sub)-Gaussianity
(according to Definition~\ref{def_scenarios} and~\eqref{eq:consistency})
and computational complexity, and 
their software implementations are listed where available.

\subsubsection{\cred{Numerical performance}}
\label{sec:empirical}

\cred{We briefly discuss the practical performance of the representative change point methods 
in Table~\ref{table:overview}, based on the numerical experiments conducted in \cite{cho2020}
using their respective implementations listed in the table;
see also \citet{fearnhead2019} for an extensive comparison of data segmentation algorithms
based on their numerical performance,
and the vignette accompanying the R package {\tt breakfast} \citep{breakfast}
comparing various CUSUM-based candidate generation and model selection methods.}

\cred{We emphasize that different algorithms excel at different performance criteria,
and their performance vary with different test signals.
For instance, dynamic programming algorithms for minimising the $\ell_0$-penalised cost functions,
such as \cite{killick2012} and \cite{maidstone2017} 
(implemented in \cite{changepoint} and \cite{fpop}, respectively),
take a fraction of a second in processing long data sequences ($n \ge 2 \times 10^4$),
followed by the multiscale MOSUM procedure with bottom-up merging \citep{chan2017, messer2014};
the SDLL model selection method of \cite{fryzlewicz2020} is well-suited to
detect a large number of frequent change points;
FDRseg \citep{li2016} shows favourable performance for relatively short test signals;
TGUH \citep{fryzlewicz2017} tends to incur few false positives 
even when change points are sparse while $n$ is large.
Between those methods that employ information criteria for model selection,
the localised pruning proposed in \cite{cho2020}
outperforms the sequential minimisation along the solution path considered in \cite{fryzlewicz2014}
in terms of correctly estimating the number of change points.
Overall, the localised pruning methodology shows well-rounded performance 
in combination of either MOSUM- or CUSUM-based candidate generating procedures
on a variety of test signals with both sparse and dense change points.}

\cred{These observations apply to the case of i.i.d.\ Gaussian errors.
Most of the methods considered above are not easily extended to cope with
heavy-tailed or serially correlated  $\{\vep_t\}$,
as their tuning parameters are specifically tailored for the i.i.d.\ (sub-)Gaussian setting.
The localised pruning, on the other hand, 
provides a framework for selecting a theoretically-motivated penalty in the SC
that is well-justified in the numerical experiments 
where $\{\vep_t\}$ are generated as i.i.d.\ random variables following $t_5$-distributions,
or AR($1$) processes with both strong and weak autocorrelations.}

\subsection{\cred{Inference for the multiple change point problem}}
\label{sec:inference}

{\color{black} In this section, we review the literature on inference for multiple change points 
in the context of the canonical segmentation problem.
For a review of the literature in the case of a single change point, see \citet{jandhyala2013inference}.

In order to quantify the uncertainty in the detection of multiple change points,
MCPS methods (see Section~\ref{sec:mcps}) such as 
SMUCE \citep{frick2014} and FDRseg \citep{li2016}
allow for controlling the probability of over-estimating the number of change points 
in terms of family-wise error rate or false discovery rate.
\citet{eichinger2018} provide the asymptotic null distribution of
the test statistic of the single-scale MOSUM procedure,
namely that of $\max_{(s, k, e) \in \mc S_{\text{M}}(G)} \vert \mc T_{s, k, e}(X)\vert$
(see~\eqref{eq:mosum}) when there is no change point present, 
under general assumptions permitting heavy tails and serial dependence in $\{\vep_t\}$.
Using this distribution, we can test whether or not there exists a mean shift
within the $G$-environment at each time point $k$,
where the thus-obtained $p$-value controls the family-wise error rate by construction.
There also exist resampling methods for change point tests in the i.i.d.\ setting
which generate better small sample approximations of the p-values, 
see \citet{antoch2001permutation}
and also \citet{huvskova2004} for a survey.
For the case of i.i.d.\ Gaussian errors, \citet{fang2020} give the approximation of the tail probability of
$\max_{0 \le s < k < e \le n} \vert \mc T_{s, k, e}(X) \vert$
(maximum of the CUSUM statistics over all possible triplets $(s, k, e)$)
under the null hypothesis of no change points,
which in turn provides an approximation of the probability of a false positive error.

Post-selection inference methods exist
for the change point problem when $\{\vep_t\}$ are i.i.d.\ Gaussian,
where the interest lies in testing for the equality of the mean
around estimated change point locations conditional on their estimation process,
see e.g.\ \citet{hyun2018post} and \citet{jewell2019testing}.
Thanks to the conditioning sets being polyhedral in the case of BS and its variants,
or via efficient characterisation of the conditioning set afforded 
by the functional pruning technique \citep{maidstone2017} 
for minimising the $\ell_0$-penalised objective functions (Section~\ref{sec:ell:zero}),
the corresponding $p$-values can be computed from truncated normal distributions;
however, their interpretation is involved and this approach requires further correction for multiple testing.
\citet{fryzlewicz2020narrowest} considers an alternative framework called post-inference selection
under which the interest lies in identifying intervals containing at least one change points at a given
global significance level without locating them.

In a more classical sense, there are also a few methods for quantifying 
the uncertainty about the locations of change points.
\citet{eichinger2018} derive the asymptotic distribution of the MOSUM-based change point estimators 
in the case of local changes (i.e.\ $d_j\to 0$),
and \citet{chokirch2021} show that the corresponding limit for the fixed change case (i.e.\ $d_j$ constant)
depends on the unknown distribution of $\{\vep_t\}$,
extending similar observations made in the AMOC setting 
for CUSUM-based estimators \citep{antoch1999estimators,antoch1995change}.
This motivates constructing bootstrap confidence intervals around change points,
as proposed by \cite{antoch1999estimators, antoch1995change} in the case of i.i.d.\ errors
and by \cite{huvskova2008, huvskova2010} in time series settings,
all for the AMOC problem.
\citet{meier2021mosum} outline the bootstrap construction of pointwise and uniform 
confidence intervals for multiple change points based on the MOSUM procedure,
and \citet{chokirch2021} show its theoretical validity for both local and fixed changes
in the presence of i.i.d.\ (not necessarily Gaussian) $\{\vep_t\}$.

In addition, MCPS \citep{frick2014,li2016} provide a confidence set for all candidate signals
from which confidence intervals around the change points can be obtained.
Using the inverse relation between confidence intervals and hypothesis tests,
\cite{fang2020} describe how approximate confidence regions can be generated.}

\section{Extensions to more complex problems}
\label{sec:ext}

In this section, we discuss several avenues for 
extending the methodologies from Section~\ref{sec:seg}
to more complex change point problems 
beyond that of detecting change points in the mean of univariate data sequence under~\eqref{eq:model},
which include high-dimensional change point analysis. 
While some literature in these directions already exists, 
it is far scarcer than the literature on the canonical data segmentation problem.
We argue that a thorough understanding 
on theoretical and computational performance of different data segmentation methods
for the simpler problem in the univariate setting,
forms the basis for the methodological development in more complex situations.

In Section~\ref{section_beyondmean},
we discuss that many tests for changes in the stochastic properties other than the mean 
(mostly in the AMOC scenario),
are closely related to those proposed for 
detecting a change in the mean of univariate data. 
Therefore, data segmentation algorithms that have proven useful for the canonical segmentation problem
can be adapted for complex change point detection problems
in combination with such tests.
These combined methodologies are expected to deliver satisfactory or even optimal performance,
although they require more sophisticated mathematical analysis
which in turn necessitates an in-depth understanding of both parts to be combined. 

In Section~\ref{section_HD}, we give a brief overview 
of the state-of-the-art methodologies for high-dimensional change point problems.
Focusing on the problem of change point detection in the mean of high-dimensional panel data,
we show that the challenges in dealing with high dimensionality
are orthogonal to those arising from the presence of multiple change points. 

\subsection{Extensions beyond detecting mean changes under Gaussianity}
\label{section_beyondmean}

First we provide a brief review of the existing literature
on the detection of multiple change points in stochastic properties other than the mean,
in univariate or multivariate data sequences.
Under parametric models for linear and non-linear time series,
$\ell_0$- \citep{aue2014, davis2006, davis2008} and
\cred{$\ell_1$-penalisation \citep{behrendt2021note, chan2014}} methods,
the MOSUM procedure based on likelihood ratio statistic \citep{yau2016}
and BS \citep{fryzlewicz2014b}
have been adopted for detecting change points in
the parameters that vary over time in a piecewise constant manner.
Data segmentation under piecewise polynomial (mostly linear) models
have been addressed in \cite{baranowski2019, fearnhead2019d, maeng2019, tibshirani2014},
also adopting the algorithms discussed in Section~\ref{sec:seg}.
For nonparametric change point analysis,
BS \citep{matteson2014}, WBS \citep{padilla2019} and
\cred{$\ell_0$-penalisation \citep{arlot2019kernel, garreau2018, zou2014}}
have been combined with various methods for measuring distributional changes,
and some computational efforts have been made 
to improve the dynamic programming algorithms 
for the methods in the latter category \citep{celisse2018, haynes2017b}.
In the time series setting, methods based on local periodograms \citep{preuss2015}
and wavelet transform \citep{cho2012} have been developed
in combination with the MOSUM procedure and BS, respectively.

As seen in the above, data segmentation problems beyond that of mean change point detection
require a methodology appropriate for revealing and measuring 
the time-varying characteristics of the data,
which is of separate concern from handling multiple change points.
We can tackle complex data segmentation problems 
by combining suitable methodologies addressing the two different aspects.
This in turn requires an in-depth understanding of both strengths and weaknesses
of the procedures for the canonical data segmentation problem as discussed in Section~\ref{sec:seg},
in addition to understanding how extensions beyond detecting mean changes are achieved.
We will use the remainder of this section to focus on the latter aspect
and provide a short overview of the techniques 
developed mainly in the AMOC setting.

The \emph{effective dimensionality} of data segmentation problems 
is given by the number of unknown parameters such that even for univariate observations, 
the effective dimensionality of a given problem can be large,
see e.g.\ \citet{kirch2012testing} for an example for testing for parameter stability
in non-linear autoregressive models by means of a semiparametric approach
based on neural networks, which involves a large number of parameters.
To make the problem even more difficult, 
statistics for revealing changes (e.g.\ obtained after a suitable data transformation) 
are often not independent even if the observed data is. 
\cred{\citet{bucher2019combining} consider a test for detecting changes in the serial dependence using a copula approach,
and discuss how the result from such a test can be combined
with those from change point tests for distributional changes
(all based on  resampling methodology) even though these tests are dependent.}
\citet{kirch2015eeg} describe in detail how testing for changes
in moderately large linear vector autoregressive processes becomes a high-dimensional problem,
and provide discussions on how to resolve the complications arising from the high dimensionality.
This shows that techniques for high-dimensional change point analysis 
as discussed later in Section~\ref{section_HD},
are also of relevance for data segmentation under complex time series models. 

\subsubsection*{Data transformation}

Some complex change point problems can be reduced to 
that of detecting changes in the mean
by adopting a suitable data transformation
where typically, the dimensionality of the transformed data 
is larger (sometimes much larger) than that of the original observation sequence.
For example, the problem of detecting changes 
in the covariance of $p$-dimensional data $\mbf X_t = (X_{1t}, \ldots, X_{pt})^\top$
(under the assumption of a constant mean),
can be transformed to that of detecting changes in the mean of 
$p (p+1)/2$-dimensional data
by considering the vector consisting of cross-products, 
i.e.\ $\mathbb{X}_t = (X_{it} X_{jt}: \, 1 \le i \le j \le p)^\top$. 
Clearly, the effective dimensionality is quadratic 
in the dimension of the original data. 
Also, the components of $\mathbb{X}_t$ are not independent 
and the dependence is not negligible.
Such a transformation has been adopted for change point analysis in the covariance structure
of multivariate \citep{aue2009break}, functional \citep{stoehr2020detecting}
and high-dimensional \citep{steland2020testing} data,
and \citet{cho2015} propose a similarly motivated wavelet transform 
for the segmentation of high-dimensional time series in the second-order structure.

\subsubsection*{Generalised method of moments}

The above approach is applicable only if the parameters of interest,
which are subject to change over time, can be expressed as the expected values
of an appropriately transformed data, i.e.\ as some moments.
For most parametric models, however, this is not the case. 
For example, suppose that $\{X_t\}$ follows a (centered) linear AR model of order one, i.e.\ 
\begin{align}
\label{eq:ar}
X_t = \alpha X_{t - 1} + e_t, \quad e_t \sim \text{WN}(0, \sigma^2), \quad t = 1, \ldots, n.
\end{align}
Then, the AR parameter $\alpha$ cannot be written as any moment after a simple data transformation.
However, it can be written as a generalised moment, namely the quotient $\E(X_t X_{t-1})/\E(X_t^2)$,
as given by the well-known Yule-Walker equation
and moreover, can be estimated (locally) from the data by using the method of moments. 
This observation gives rise to the following two approaches for change point analysis.

The Wald approach compares directly the distance between the estimators 
obtained from different stretches of the data: At location $k$,
we evaluate $d(\widehat{\beta}_{0:k}, \widehat{\beta}_{k:n})$
where $d$ is some measure of distance, and
$\widehat{\beta}_{a:b}$ denotes the estimator for the parameter of interest $\beta$ 
obtained from the data stretch $\{X_t, \, a < t \le b\}$.
The Wald approach leads to good performance 
if the parameters are identifiable and 
the local estimators can be obtained without resorting to numerical algorithms. 
On the other hand, its application may not be appropriate
when there exist many (local) optima 
of the objective function for parameter estimation.
Such a situation arises e.g.\ for neural networks,
where different parameters describing the same or very similar regression functions
might be far from each other in the parameter space. 
Further, this is particularly problematic if the estimators can only be obtained numerically.
Even when the objective function does not have multiple local maxima, 
if it is relatively flat around the true parameter,
it can be misleading to gauge the distance between the estimators of the parameters 
instead of the processes that they describe.

A score-type approach is more robust in this respect and 
is closely related to the CUSUM statistic described in Section~\ref{sec:loc}.
In contrast to simple data transformations as described previously,
the transformations adopted in the score-type approach
often involve the (local) parameter estimators. 
Let us illustrate this by re-visiting the example of a linear AR(1) time series~\eqref{eq:ar}
in the AMOC situation, where the score-type statistic is given by
\begin{align}
\label{eq:score}
\frac{1}{\sqrt{n}} \max_{2 \le k < n}\left|\sum_{t = 2}^k X_{t - 1} (X_t - \wh{\alpha}_{0:n} X_{t-1}) \right|,
\end{align}
where $\wh{\alpha}_{0:n}$ denotes an estimator of the AR parameter based on the \emph{full} data set. 
Some authors suggest to use partial sums of estimated residuals; 
however, the thus-constructed test has power only against some alternatives,
namely those leading to changes in the mean of estimated residuals (see e.g.\ \citep{kirch2015eeg}). 
The statistic in~\eqref{eq:score} bears clear resemblance to the CUSUM statistic
for mean change detection 
(written below with slightly different location-dependent weights from Section~\ref{sec:loc}
for better comparison):
\begin{align*}
\frac{1}{\sqrt{n}} \max_{1 \le k < n} \left| \sum_{t = 1}^k (X_t - \bar{X}_{0:n})\right|.
\end{align*}
Here, $\bar{X}_{0:n}$ can be regarded as an estimator of the unknown mean based on the full sample, 
analogously to $\wh{\alpha}_{0:n}$ being the global estimator of $\alpha$. 

There are many other examples where the score-type statistics are adopted for change point analysis, 
including the tests based on likelihood ratio or classical $M$-estimators 
\cred({see the next subsection on robust change point analysis)}
and those for changes in linear (auto)regression \citep{huvskova2007detection, kirch2015eeg}, 
integer-valued time series such as Poisson autoregressive processes \citep{kirch2014detection} 
and even regression models based on neural networks \citep{kirch2012testing}. 
In all of the above, the test statistics are based on partial sums 
\begin{align*}
\frac{1}{\sqrt{n}} \max_{1 \le k < n} \left\| \sum_{t = 1}^k F(\mathbf{Y}_t, \widehat{\beta}_{0:n}) \right\|
\quad \text{where} \quad \sum_{t = 1}^n F(\mathbf{Y}_t, \widehat{\beta}_{0:n}) = 0,
\end{align*}
where $F$ is a suitable estimating function and 
$\widehat{\beta}_{0:n}$ the global estimator based on the full sample from this estimating function. 
The norm $\|\cdot\|$ often involves the unknown covariance matrix of $F(\mathbf{Y}_t, \beta)$ 
or an appropriate estimator thereof.
In many instances, $\mathbf{Y}_t$ is not the original observation sequence;
e.g.\ in the above AR($1$) example, 
$\mathbf{Y}_t$ contains lagged observations.
This shows that effectively, one is looking for mean changes in the pseudo-data 
$F(\mathbf{Y}_t, \widehat{\beta}_{0:n})$, 
whose dimension is equal to the number of unknown parameters. 
For linear models, the Wald and the score-type statistics are equivalent in the AMOC setting,
see \citet{kirch2014detection}. 
In non-linear situations, this does not hold in small samples but does so asymptotically 
in well-behaved situations. 

The Wald approach can be extended to multiple change point detection
e.g.\ in combination with the MOSUM procedure
by comparing the distance between local estimators over moving windows.
Alternatively, the corresponding moving sum score statistic with a local estimator can be used.
However, both suffer from the problem of having to calculate $O(n)$ local estimators
which, as noted above, brings in further numerical challenges
when there does not exist an analytic solution to the estimating function.
Using a slightly modified score-type statistic
replacing the local estimator with
a global \emph{inspection parameter}, say $\wt{\beta}$ (which may be data-dependent), 
alleviates this issue but typically, there is no guarantee that all changes are detectable as changes 
in the mean of $F(\mathbf{Y}_t, \wt{\beta})$.
\citet{reckruhm2019} shows that even so, at least one change is made detectable therein,
which naturally leads to adopting several of such inspection parameters.
Some first results have been obtained by 
\citet{reckruhm2019} and \citet{kirchreckruehm2020},
on the combined methodology of the MOSUM procedure and the use of estimating functions
for the detection of multiple change points in general settings.

\subsubsection*{\cred{Robust change point analysis}}

{\color{black} The framework of generalised method of moments also prominently includes 
robust tests for changes in the mean or regression parameters
(often known as $M$-estimation in this context). 
Selecting a suitable estimating function $F$ (related to what is known as a score function),
one can obtain not only more robust estimators of the parameters of interest 
but also more robust methods to test for their changes.
For example, under the mean change model as in~\eqref{eq:model} with $q_n \le 1$,
choosing $F(X, \beta) = \bbI_{\{X >\beta\}} - \bbI_{\{X < \beta\}}$ 
(with $\beta$ corresponding to $f_0 = \E(X_t)$ when there exists no change point) results in 
$\wh{\beta}_{0:n}$ that coincides with the median 
where $\sum_{t = 1}^n F(X_t, \wh{\beta}_{0:n}) = 0$. 
Such an approach is also known as $L_1$-procedure and discussed 
e.g.\ in \citet{huvskova1990asymptotics} for detecting changes in linear regression.
Theoretical investigation into change point detection procedures based on $M$-estimation
can be found in \cite{antoch1994procedures, huvskova1990asymptotics, huvskova1990some, huvskova1996tests, huvskova2002m} under the i.i.d.\ assumption,  
and in \cite{huvskova2012m, pravskova2014m} for dependent data.
A more detailed survey of the literature on this topic, 
including references on a-posteriori as well as sequential change point testing,
is given in \cite{huvskova2013robust}.
The paper includes some discussions on rank-based tests as well,
which is an alternative approach to achieve robustness in change point testing.
One such an example is the generalisation of the Wilcoxon rank sum test  to the change point setting
which is a special case of a (non-degenerate) $U$-statistic. 
$U$-statistics are adopted for the estimation of a range of different parameters far beyond the mean, 
and also used in testing for the stability of the corresponding parameters.
Importantly, they involve a data transformation that translates the problem at hand
into that of mean change point testing.
To elaborate, the Hoeffding decomposition 
is an important tool to prove asymptotic results for $U$-statistics
and shows equality between the original statistic and 
a sum of i.i.d.\ random variables (obtained from the Haj\'ek projection) 
plus an asymptotically negligible remainder.
Change point tests based on $U$-statistics were first proposed in the i.i.d.\ setting
\citep{csorgHo1989invariance, ferger1994power, gombay2001u}
and then generalised to weakly \citep{dehling2015change} and even long-range dependent data \citep{betken2016testing,dehling2013non}.
Extensions based on $U$-quantiles are also available \citep{dehling2020robust, vogel2017studentized}.
Some first results for the data segmentation problem based on $U$-statistics already exist  
\citep{DORING20102003, orasch2004using},
and \citet{fearnhead2019} investigate the minimisation of penalised costs 
based on $M$-estimation via dynamic programming for change point analysis in the presence of outliers.}

\subsubsection*{Asymptotic embeddings}

Another extension is based on asymptotic results for partial sum processes 
such as \cred{functional central limit theorems or (strong) invariance principles
\citep{berkes2014, gorecki2018change, Heunis, komlos1975, komlos1976, kuelbs1980, SteinebachEastwood}}.
Most importantly, such results enable the investigation into change point analysis of stochastic processes  
including multivariate Wiener processes with drift, renewal processes or integrated diffusion processes.
A univariate version of such a model was introduced by \citet{horvath2000testing} 
for change point testing against the AMOC alternative,
and has been extended to multiple change point problems in \citet{kuhn2001estimator}
where the consistency is established for the estimator of the number of change points 
based on the SC described in Section~\ref{sec:ell:zero}.
For the specific example of univariate renewal processes, 
\citet{messer2018} propose a multiscale MOSUM procedure with bandwidths linear in sample size
(see Sections~\ref{sec:mosum} and~\ref{sec:ms}).
\citet{klein2020} investigate a MOSUM procedure in the general multivariate framework
adopting both linear and sub-linear bandwidths
and derive localisation rates for the change point estimators
that are shown to be minimax optimal 
when there are a finite number of change points; 
when there are an unbounded number of change points, 
this remains to be valid for Wiener processes with drift.
Beyond such settings, minimax optimality in the localisation rate
has not been established to the best of our knowledge (see also Section~\ref{sec:cp}).

\subsubsection*{\cred{Nonparametric change point analysis}}

Truly nonparametric change point tests do not have obvious connections 
to the canonical mean change point detection problem.
These tests often involve appropriately weighted integrals over
differences of the estimators of the nonparametric quantity of interest,
such as those based on functionals of two-sample rank-based test statistics \citep{yao1990asymptotic},
empirical characteristic function \citep{huvskova2006change, huvskova2006change1},
and kernel estimators of nonparametric conditional mean functions 
\citep{mohr2020consistent} among others.
In principle, local test methods for the canonical data segmentation problem 
discussed in Section~\ref{sec:loc},
such as (W)BS or the MOSUM procedure, can be adapted to accommodate such tests,
since they are based on localised applications of tests designed for the AMOC setting. 
Application of the global optimisation approaches 
to nonparametric settings is more involved, although some attempts exist \citep{zou2014}.

\subsection{High-dimensional change point analysis}
\label{section_HD}

Recent years have seen a surge of interest in the development of change point methodologies
for high-dimensional data including functional and panel data and networks,
with many papers addressing the problem of change point testing
in the mean of panel data,
see e.g.\ Sections~7 and~8 of the review paper by \citet{horvath2014}.
In this section, our aim is to show that the challenge due to high dimensionality of the data
is orthogonal to that due to the presence of multiple change points
in the following sense:
\begin{enumerate}[label = (\roman*)]
\item \label{hd:one} As argued in Section~\ref{sec:cp}, the goal of the canonical data segmentation problem
is to achieve consistency in multiple change point detection
even in difficult situations arising from the multiscale nature of the change points
(see Definition~\ref{def_scenarios}~\ref{def:dlb}). 
In particular, the interplay between the spacing between change points,
$\delta_j = \min(\cp_j - \cp_{j - 1}, \cp_{j + 1} - \cp_j)$, and
the magnitude of changes, $\vert d_j \vert$, in their detectability is well-understood under~\eqref{eq:model}. 
Accordingly, the methodological challenge lies in 
isolating each change point within an interval that is
long enough for its detection and localisation.

\item \label{hd:two} In high-dimensional change point problems, 
the main question is how to boost the signal (on each change) across large dimensions. 
Adopting a minimax-viewpoint, this is closely related to the choice of the norm 
adopted for measuring the cross-sectional size of the change,
which is highly relevant in high-dimensional settings since,
unlike in univariate or even multivariate situations,
standard norms are no longer equivalent. 
\end{enumerate}

Clearly, these two aspects,
multiple change point detection and high-dimensional change point analysis, 
merit separate attention when developing a methodology 
that is most suitable for the problem at hand. 
There already exist papers addressing 
the problem of multiple change point detection in high-dimensional settings
through combining the data segmentation algorithms for the canonical data segmentation problem
with methods for cross-sectional aggregation of information on changes.
We provide an incomplete review of the literature below.

The application of the BS algorithm has been investigated in panel data segmentation problems
based on first- \citep{cho2016} and second-order \citep{cho2015} properties, 
as well as under a factor model \citep{barigozzi2018}.
The WBS algorithm has been adopted for
multiple change point detection in the mean \citep{wang2018} and
the covariance \citep{wang2017} of high-dimensional panel data
in combination with data-driven projections,
and also adopted in regression settings \citep{wang2019b}.
\citet{chu2019} and \citet{liu2020fast} propose graph-based change point tests 
for non-Euclidean and possibly high-dimensional data, 
and the latter paper heuristically discusses their extensions to data segmentation problems
in combination with BS and WBS.
\citet{chen2019} adopt the MOSUM procedure for panel data segmentation in the mean,
and a MOSUM-type screening strategy has also been adopted in \citet{zhao2019} 
for change point detection in dynamic networks.
The $\ell_0$-penalisation method has been adopted in parametric settings
such as regression \citep{kaul2019, leonardi2016} and vector autoregressive \citep{wang2019} models
and the $\ell_1$-penalisation method, together with an information criterion-based model selection,
has been applied to high-dimensional time series segmentation under vector autoregressive models
\citep{bai2020, safikhani2020}.

We further elaborate on the issue discussed in \ref{hd:two} above
by focusing on the problem that has seen the most development, 
namely the problem of detecting changes in the mean of panel data
under the following model:
\begin{align}
\label{eq:model:hd}
X_{it} = f_{it} + \vep_{it} = f_{i0} + \sum_{j = 1}^{q_n} d_{ij} \cdot \mathbb{I}_{t \ge \cp_j + 1} + \vep_{it}, 
\quad  i = 1, \ldots, p, \, t = 1, \ldots, n.
\end{align}
It extends the canonical model~\eqref{eq:model} to the panel data setting
with the vector $\mbf d_j = (d_{ij}, \ldots, d_p)^\top$ denoting the change in the mean of 
$\mbf X_t = (X_{1t}, \ldots, X_{pt})^\top$ at $t = \cp_j$,
and $\bm\vep_t = (\vep_{1t}, \ldots, \vep_{pt})^\top$
is assumed to satisfy $\E(\bm\vep_t) = \mbf 0$ and $\Cov(\bm\vep_t) = \bm\Sigma$.
Unlike in the multivariate setting, the dimensionality of the data $p = p_n$
depends on the sample size $n$ and is even allowed to increase faster than $n$.

Detectability of a change point depends on the magnitude of the corresponding change, 
and selecting a norm for measuring the magnitude of $\mbf d_j$ 
effectively corresponds to modelling the (cross-sectional) type of a change
for which the proposed change point methodology is particularly tailored.
High-dimensional efficiency of change point tests, as proposed in \citet{aston2018high},
relates to the separation rate between detectable and undetectable changes.
The norms associated with the high-dimensional efficiency of
different methods proposed in the existing literature indeed vary:
Methods designed for detecting sparse changes 
(where $\Vert \mbf d_j \Vert_0 := \sum_{i = 1}^p \mathbb{I}_{\vert d_{ij} \vert > 0} \ll p$)
such as those proposed in \citet{jirak2014}, \citet{cho2015}, \citet{wang2018}, \citet{chen2019}
and the scan statistic of \citet{enikeeva2019},
are associated with the $\ell_\infty$-norm
(sometimes implicitly so via $\Vert \mbf d_j \Vert/\sqrt{\Vert \mbf d_j \Vert_0}$),
while the methods from \citet{horvath2012}, \citet{wvs2019}
and the linear statistic of \citet{enikeeva2019}
are associated with the $\ell_2$-norm, 
see Table~\ref{table:hde} for a summary which extends Table~1 of \citet{cho2016}.

\begin{table}[htb]
\caption{Comparison of high-dimensional change point detection methods
for the problem~\eqref{eq:model:hd}
on their high-dimensional efficiency \citep{aston2018high} when $q_n = 1$,
$\min(\cp_1, n - \cp_1) \ge cn$ for some $c \in (0, 1/2]$ and 
$\bm\Sigma$ is diagonal with all the variances of the same order, i.e.\
$0 < \sigma^2_{\min} \le \Var(\vep_{it}) \le \sigma^2_{\max} < \infty$ for all $i = 1, \ldots, p$.
The efficiency, denoted by $\mc E(\mbf d_1)$,
indicates that the corresponding procedure has asymptotic power one
if $\sqrt{n} \mc E(\mbf d_1) \to \infty$.
By $L_{n, p}$, we denote a term logarithmic in $n$ and/or $p$
which varies from one instance to another.}
\label{table:hde}
\centering
\begin{tabular}{l | c || l | c}
\hline\hline
\cite{aston2018high}: oracle & ${\Vert \mbf d_1 \Vert}$ &
\cite{aston2018high}: random & ${ p^{-1/2} \Vert \mbf d_1 \Vert}$
\\ \hline
\cite{horvath2012, wvs2019} & ${ p^{-1/4} \Vert \mbf d_1 \Vert}$ &
\cite{cho2015, jirak2014, chen2019} & ${L_{n, p}^{-1} \Vert \mbf d_1 \Vert_\infty}$
\\ \hline
\cite{enikeeva2019}: linear & ${(p^{1/4} L_{n, p})^{-1} \Vert \mbf d_1 \Vert}$ &
\cite{enikeeva2019}: scan, \cite{wang2018} & ${(\Vert \mbf d_1 \Vert_0 L_{n, p})^{-1/2} \Vert \mbf d_1 \Vert}$ 
\\ \hline
\end{tabular}
\end{table}

Minimax optimality is also linked to the choice of norm, in the sense that
minimax optimality of a given procedure with respect to a norm 
indicates that the procedure is optimal in the worst possible case
defined by that particular norm.
On the other hand, this does not provide much information as to
how the procedure performs in the worst case scenarios defined with different norms.
For example, the test proposed by \citet{horvath2012} is minimax optimal
with respect to the $\ell_2$-norm,
when the panel is spatially independent and all variances are of the same order
(in the sense described in Table~\ref{table:hde}),
see the discussion below Theorem~2.1 of \cite{aston2018high}.
The test statistic proposed therein
increases the signal-to-noise ratio (while controlling for the size)
by summing up squared change point statistics;
as more cross-sectional components are considered,
the signal increases according to the squared $\ell_2$-norm,
i.e.\ linear in the number of components with a change,
whereas the noise only increases as $\sqrt{p}$
due to the central limit theorem applicable under the spatial independence assumption.
However, it does not adapt to the sparsity of change
or match the minimax optimal rate with respect to the $\ell_\infty$-norm;
we refer to \citet{liu2019} where the authors show the presence of a phase transition boundary for the sparsity
in the minimax rate of the high-dimensional change testing problem under~\eqref{eq:model:hd}
(i.e.\ $q_n \le 1$).

The difficulty in high-dimensional data analysis is 
that spurious low-dimensional signals appear simply by chance, 
another flavour of the curse of dimensionality
which needs to be accounted for in any data science problem. 
Under~\eqref{eq:model:hd} with a single change point, 
the direction of a change, $\mbf d_1$, is a one-dimensional object
such that the projection of $\mbf X_t$
with respect to the `oracle' projection vector $\mbf p_{\text{o}} = \bm\Sigma^{-1} \mbf d_1$
yields the optimal signal-to-noise ratio, see Proposition~3.3 of \cite{aston2018high}.
In practice, however, neither $\mbf d_1$ nor $\bm\Sigma$ are known,
and maximising over all possible projections $\mbf p \in \R^p$ 
will lead to too large a noise level due to the above observation.
Random projections (independent of the data) have been considered as an alternative,
see \citet{steland2020testing} addressing the problem of change point testing in the covariance structure.
However, this is at the cost of the loss of detection power 
as illustrated by the high-dimensional efficiency 
of such random projections, see Table~\ref{table:hde}
which shows that a random projection is a magnitude of $p^{-1/2}$ worse than oracle efficiency
when $\bm\Sigma$ is a diagonal matrix, and also Theorem~3.5 of \cite{aston2018high}.

Data-driven projections such as those proposed in 
\cite{chen2019, cho2015, enikeeva2019, wang2018}
can keep the increased noise level at bay, 
by effectively working under the sparsity assumption on $\mbf d_j$
and consequently considering only a subset of all possible projections. 
This comes at the cost of decreased high-dimensional efficiency 
when the sparsity assumption is not met,
compared to the oracle projection with respect to $\mbf p_{\text{o}}$
or the $\ell_2$-norm minimax optimal procedures such as \citep{horvath2012, wvs2019}.

Compared to the panel data setting,
change point analysis in functional data is benefited by
the additional structure implied by the assumption of observing a random function,
which helps lifting part of the curse of dimensionality.
Some approaches \citep{aston2012detecting, berkes2009detecting} 
to the problem of detecting changes in the mean function 
are based on dimension reduction e.g.\ via functional principal component analysis. 
Another line of research proposes 
`fully functional' procedures \citep{aue2018detecting, sharipov2016sequential}
in the sense that they do not employ any dimension reduction. 
\citet{stoehr2020detecting} propose a compromise between the two approaches
in the context of change point detection in the covariance structure.

\section{\cred{Conclusions}}
\label{sec:conc}

{\color{black} In this paper, we first review and compare different approaches 
to the canonical segmentation problem
where the aim is to estimate the number and the locations of multiple change points
in the mean of univariate data. 
In Section~\ref{sec:overview}, we provide an overview of representative methodologies
comparing them according to the theoretical framework given in Section~\ref{sec:cp},
and briefly comment on their numerical performance
based on comparative simulation studies on a range of test signals.

In the second part of the paper, we give a survey on topics
such as change point analysis beyond the detection of mean changes under Gaussianity 
(Section~\ref{section_beyondmean}),
and high-dimensional change point analysis (Section~\ref{section_HD}).
In Section~\ref{section_beyondmean}, we argue that 
change point tests developed for more complex problems 
are closely related to the mean change point problem
and make the connection transparent. 
As discussed in Section~\ref{sec:cp}, some of the methods for the canonical data segmentation problem
are based on localised applications of the tests for the AMOC alternative, 
and thus can be applied to more complex change point problems 
in combination with corresponding tests.
Clearly, theoretical underpinnings of the combined methodology
depend on those of the individual components.
In Section~\ref{section_HD}, we show that in high-dimensional change point analysis,
challenges arising from the high dimensionality are orthogonal to those arising 
from the presence of multiple change points and,
based on in-depth understanding of the component problems,
methods addressing individual aspects can be combined to best address the given problems.
Finally, combining the elements from the three fields opens up a new avenue for future research, 
namely the development of statistical methodologies 
for complex, high-dimensional data segmentation problems.}

{\small
\singlespacing
\setlength{\bibsep}{0.0pt}
\bibliographystyle{plainnat}
\bibliography{fbib}

\begin{thebibliography}{196}
\providecommand{\natexlab}[1]{#1}
\providecommand{\url}[1]{\texttt{#1}}
\expandafter\ifx\csname urlstyle\endcsname\relax
  \providecommand{\doi}[1]{doi: #1}\else
  \providecommand{\doi}{doi: \begingroup \urlstyle{rm}\Url}\fi

\bibitem[Adams and Heard(2016)]{adams2016}
Niall~M Adams and Nicholas Heard.
\newblock \emph{Dynamic Networks and Cyber-security}, volume~1.
\newblock World Scientific, 2016.

\bibitem[Aminikhanghahi and Cook(2017)]{aminikhanghahi2017}
Samaneh Aminikhanghahi and Diane~J Cook.
\newblock A survey of methods for time series change point detection.
\newblock \emph{Knowledge and Information Systems}, 51:\penalty0 339--367,
  2017.

\bibitem[Anastasiou and Fryzlewicz(2018)]{IDetect}
Andreas Anastasiou and Piotr Fryzlewicz.
\newblock \emph{IDetect: Detecting multiple generalized change-points by
  isolating single ones}, 2018.
\newblock URL \url{https://CRAN.R-project.org/package=IDetect}.
\newblock {R} package version 1.0.

\bibitem[Anastasiou and Fryzlewicz(2021)]{anastasiou2019}
Andreas Anastasiou and Piotr Fryzlewicz.
\newblock Detecting multiple generalized change-points by isolating single
  ones.
\newblock \emph{Metrika}, pages 1--34, 2021.

\bibitem[Anastasiou et~al.(2021)Anastasiou, Chen, Cho, and
  Fryzlewicz]{breakfast}
Andreas Anastasiou, Yining Chen, Haeran Cho, and Piotr Fryzlewicz.
\newblock \emph{breakfast: Methods for Fast Multiple Change-Point Detection and
  Estimation}, 2021.
\newblock URL \url{https://CRAN.R-project.org/package=breakfast}.
\newblock {R} package version 2.2.

\bibitem[Antoch and Hu{\v{s}}kov{\'a}(1994)]{antoch1994procedures}
Jarom{\'\i}r Antoch and Marie Hu{\v{s}}kov{\'a}.
\newblock Procedures for the detection of multiple changes in series of
  independent observations.
\newblock In \emph{Asymptotic Statistics}, pages 3--20. Springer, 1994.

\bibitem[Antoch and Hu{\v{s}}kov{\'a}(1999)]{antoch1999estimators}
Jaromir Antoch and Marie Hu{\v{s}}kov{\'a}.
\newblock Estimators of changes.
\newblock In \emph{Asymptotics, Nonparametrics, and Time Series}, pages
  557--561. CRC Press, 1999.

\bibitem[Antoch and Hu{\v{s}}kov{\'a}(2001)]{antoch2001permutation}
Jarom{\i}r Antoch and Marie Hu{\v{s}}kov{\'a}.
\newblock Permutation tests in change point analysis.
\newblock \emph{Statistics \& Probability Letters}, 53\penalty0 (1):\penalty0
  37--46, 2001.

\bibitem[Antoch et~al.(1995)Antoch, Hu{\v{s}}kov{\'a}, and
  Veraverbeke]{antoch1995change}
Jaromir Antoch, Marie Hu{\v{s}}kov{\'a}, and No{\"e}l Veraverbeke.
\newblock Change-point problem and bootstrap.
\newblock \emph{Journal of Nonparametric Statistics}, 5\penalty0 (2):\penalty0
  123--144, 1995.

\bibitem[Arias-Castro et~al.(2011)Arias-Castro, Candes, and Durand]{arias2011}
Ery Arias-Castro, Emmanuel~J Candes, and Arnaud Durand.
\newblock Detection of an anomalous cluster in a network.
\newblock \emph{The Annals of Statistics}, 39:\penalty0 278--304, 2011.

\bibitem[Arlot and Celisse(2011)]{arlot2011segmentation}
Sylvain Arlot and Alain Celisse.
\newblock Segmentation of the mean of heteroscedastic data via
  cross-validation.
\newblock \emph{Statistics and Computing}, 21\penalty0 (4):\penalty0 613--632,
  2011.

\bibitem[Arlot et~al.(2019)Arlot, Celisse, and Harchaoui]{arlot2019kernel}
Sylvain Arlot, Alain Celisse, and Zaid Harchaoui.
\newblock A kernel multiple change-point algorithm via model selection.
\newblock \emph{Journal of Machine Learning Research}, 20\penalty0 (162), 2019.

\bibitem[Aston and Kirch(2012)]{aston2012detecting}
John~AD Aston and Claudia Kirch.
\newblock Detecting and estimating changes in dependent functional data.
\newblock \emph{Journal of Multivariate Analysis}, 109:\penalty0 204--220,
  2012.

\bibitem[Aston and Kirch(2018)]{aston2018high}
John~AD Aston and Claudia Kirch.
\newblock High dimensional efficiency with applications to change point tests.
\newblock \emph{Electronic Journal of Statistics}, 12:\penalty0 1901--1947,
  2018.

\bibitem[Aue and Horv{\'a}th(2013)]{aue2013}
Alexander Aue and Lajos Horv{\'a}th.
\newblock Structural breaks in time series.
\newblock \emph{Journal of Time Series Analysis}, 34:\penalty0 1--16, 2013.

\bibitem[Aue et~al.(2009)Aue, H{\"o}rmann, Horv{\'a}th, and
  Reimherr]{aue2009break}
Alexander Aue, Siegfried H{\"o}rmann, Lajos Horv{\'a}th, and Matthew Reimherr.
\newblock Break detection in the covariance structure of multivariate time
  series models.
\newblock \emph{The Annals of Statistics}, 37:\penalty0 4046--4087, 2009.

\bibitem[Aue et~al.(2014)Aue, Cheung, Lee, and Zhong]{aue2014}
Alexander Aue, Rex~CY Cheung, Thomas~CM Lee, and Ming Zhong.
\newblock Segmented model selection in quantile regression using the minimum
  description length principle.
\newblock \emph{Journal of the American Statistical Association}, 109:\penalty0
  1241--1256, 2014.

\bibitem[Aue et~al.(2018)Aue, Rice, and S{\"o}nmez]{aue2018detecting}
Alexander Aue, Gregory Rice, and Ozan S{\"o}nmez.
\newblock Detecting and dating structural breaks in functional data without
  dimension reduction.
\newblock \emph{Journal of the Royal Statistical Society, Series B},
  80:\penalty0 509--529, 2018.

\bibitem[Auger and Lawrence(1989)]{auger1989}
Ivan~E Auger and Charles~E Lawrence.
\newblock Algorithms for the optimal identification of segment neighborhoods.
\newblock \emph{Bulletin of Mathematical Biology}, 51:\penalty0 39--54, 1989.

\bibitem[Axt and Fried(2020)]{axt2020}
Ieva Axt and Roland Fried.
\newblock On variance estimation under shifts in the mean.
\newblock \emph{ASta Advanced in Statistical Analysis}, 104:\penalty0 417--457,
  2020.

\bibitem[Bai and Perron(1998)]{bai1998estimating}
Jushan Bai and Pierre Perron.
\newblock Estimating and testing linear models with multiple structural
  changes.
\newblock \emph{Econometrica}, pages 47--78, 1998.

\bibitem[Bai et~al.(2020)Bai, Safikhani, and Michailidis]{bai2020}
Peiliang Bai, Abolfazl Safikhani, and George Michailidis.
\newblock Multiple change points detection in low rank and sparse high
  dimensional vector autoregressive models.
\newblock \emph{IEEE Transactions on Signal Processing}, 68:\penalty0
  3074--3089, 2020.

\bibitem[Baranowski et~al.(2019)Baranowski, Chen, and
  Fryzlewicz]{baranowski2019}
Rafal Baranowski, Yining Chen, and Piotr Fryzlewicz.
\newblock Narrowest-over-threshold detection of multiple change-points and
  change-point-like features.
\newblock \emph{Journal of the Royal Statistical Society, Series B},
  81:\penalty0 649--672, 2019.

\bibitem[Barigozzi et~al.(2018)Barigozzi, Cho, and Fryzlewicz]{barigozzi2018}
Matteo Barigozzi, Haeran Cho, and Piotr Fryzlewicz.
\newblock Simultaneous multiple change-point and factor analysis for
  high-dimensional time series.
\newblock \emph{Journal of Econometrics}, 206:\penalty0 187--225, 2018.

\bibitem[Barry and Hartigan(1993)]{barry1993}
Daniel Barry and John~A Hartigan.
\newblock A {Bayesian} analysis for change point problems.
\newblock \emph{Journal of the American Statistical Association}, 88:\penalty0
  309--319, 1993.

\bibitem[Behrendt and Schweikert(2021)]{behrendt2021note}
Simon Behrendt and Karsten Schweikert.
\newblock A note on adaptive group {Lasso} for structural break time series.
\newblock \emph{Econometrics and Statistics}, 17:\penalty0 156--172, 2021.

\bibitem[Berkes et~al.(2009)Berkes, Gabrys, Horv{\'a}th, and
  Kokoszka]{berkes2009detecting}
Istv{\'a}n Berkes, Robertas Gabrys, Lajos Horv{\'a}th, and Piotr Kokoszka.
\newblock Detecting changes in the mean of functional observations.
\newblock \emph{Journal of the Royal Statistical Society, Series B},
  71:\penalty0 927--946, 2009.

\bibitem[Berkes et~al.(2014)Berkes, Liu, and Wu]{berkes2014}
Istv{\'a}n Berkes, Weidong Liu, and Wei~Biao Wu.
\newblock {Koml{\'o}s--Major--Tusn{\'a}dy approximation under dependence}.
\newblock \emph{The Annals of Probability}, 42:\penalty0 794--817, 2014.

\bibitem[Betken(2016)]{betken2016testing}
Annika Betken.
\newblock Testing for change-points in long-range dependent time series by
  means of a self-normalized wilcoxon test.
\newblock \emph{Journal of Time Series Analysis}, 37\penalty0 (6):\penalty0
  785--809, 2016.

\bibitem[Boysen et~al.(2009)Boysen, Kempe, Liebscher, Munk, and
  Wittich]{boysen2009}
Leif Boysen, Angela Kempe, Volkmar Liebscher, Axel Munk, and Olaf Wittich.
\newblock Consistencies and rates of convergence of jump-penalized least
  squares estimators.
\newblock \emph{The Annals of Statistics}, 37:\penalty0 157--183, 2009.

\bibitem[Brodsky and Darkhovsky(2000)]{brodsky2000}
B.~E. Brodsky and B.~S. Darkhovsky.
\newblock \emph{Nonparametric Statistical Diagnosis: Problems and Methods},
  volume 509.
\newblock Springer Science \& Business Media, 2000.

\bibitem[B{\"u}cher et~al.(2019)B{\"u}cher, Fermanian, and
  Kojadinovic]{bucher2019combining}
Axel B{\"u}cher, Jean-David Fermanian, and Ivan Kojadinovic.
\newblock Combining cumulative sum change-point detection tests for assessing
  the stationarity of univariate time series.
\newblock \emph{Journal of Time Series Analysis}, 40\penalty0 (1):\penalty0
  124--150, 2019.

\bibitem[Celisse et~al.(2018)Celisse, Marot, Pierre-Jean, and
  Rigaill]{celisse2018}
Alain Celisse, Guillemette Marot, Morgane Pierre-Jean, and GJ~Rigaill.
\newblock New efficient algorithms for multiple change-point detection with
  reproducing kernels.
\newblock \emph{Computational Statistics \& Data Analysis}, 128:\penalty0
  200--220, 2018.

\bibitem[Chakar et~al.(2017)Chakar, Lebarbier, L{\'e}vy-Leduc, and
  Robin]{chakar2017robust}
Souhil Chakar, E~Lebarbier, C{\'e}line L{\'e}vy-Leduc, and St{\'e}phane Robin.
\newblock {A robust approach for estimating change-points in the mean of an
  AR(1) process}.
\newblock \emph{Bernoulli}, 23\penalty0 (2):\penalty0 1408--1447, 2017.

\bibitem[Chan and Chen(2017)]{chan2017}
Hock~Peng Chan and Hao Chen.
\newblock Multi-sequence segmentation via score and higher-criticism tests.
\newblock \emph{arXiv preprint, arXiv:1706.07586}, 2017.

\bibitem[Chan et~al.(2014)Chan, Yau, and Zhang]{chan2014}
Ngai~Hang Chan, Chun~Yip Yau, and Rong-Mao Zhang.
\newblock Group {Lasso} for structural break time series.
\newblock \emph{Journal of the American Statistical Association}, 109:\penalty0
  590--599, 2014.

\bibitem[Chen et~al.(2006)Chen, Gupta, and Pan]{chen2006}
Jiahua Chen, AK~Gupta, and Jianmin Pan.
\newblock Information criterion and change point problem for regular models.
\newblock \emph{Sankhy{\=a}: The Indian Journal of Statistics}, 68:\penalty0
  252--282, 2006.

\bibitem[Chen and Gupta(2011)]{chen2011parametric}
Jie Chen and Arjun~K Gupta.
\newblock \emph{Parametric statistical change point analysis: with applications
  to genetics, medicine, and finance}.
\newblock Springer Science \& Business Media, 2nd edition, 2011.

\bibitem[Chen et~al.(2021)Chen, Wang, and Wu]{chen2019}
Likai Chen, Weining Wang, and Weibiao Wu.
\newblock Inference of break-points in high-dimensional time series.
\newblock \emph{Journal of the American Statistical Association (to appear)},
  2021.

\bibitem[Cho(2016)]{cho2016}
Haeran Cho.
\newblock Change-point detection in panel data via double {CUSUM} statistic.
\newblock \emph{Electronic Journal of Statistics}, 10:\penalty0 2000--2038,
  2016.

\bibitem[Cho and Fryzlewicz(2012)]{cho2012}
Haeran Cho and Piotr Fryzlewicz.
\newblock Multiscale and multilevel technique for consistent segmentation of
  nonstationary time series.
\newblock \emph{Statistica Sinica}, 22:\penalty0 207--229, 2012.

\bibitem[Cho and Fryzlewicz(2015{\natexlab{a}})]{cho2015}
Haeran Cho and Piotr Fryzlewicz.
\newblock Multiple change-point detection for high-dimensional time series via
  sparsified binary segmentation.
\newblock \emph{Journal of the Royal Statistical Society, Series B},
  77:\penalty0 475--507, 2015{\natexlab{a}}.

\bibitem[Cho and Fryzlewicz(2015{\natexlab{b}})]{cho2015c}
Haeran Cho and Piotr Fryzlewicz.
\newblock Corrections on `multiple change-point detection for high-dimensional
  time series via sparsified binary segmentation'.
\newblock
  \url{https://people.maths.bris.ac.uk/~mahrc/papers/sbs_correction.pdf},
  2015{\natexlab{b}}.

\bibitem[Cho and Fryzlewicz(2020)]{cho2020b}
Haeran Cho and Piotr Fryzlewicz.
\newblock {Multiple change point detection under serial dependence: Wild energy
  maximisation and gappy Schwarz criterion}.
\newblock \emph{arXiv preprint arXiv:2011.13884}, 2020.

\bibitem[Cho and Kirch(2020)]{cho2020}
Haeran Cho and Claudia Kirch.
\newblock {Two-stage data segmentation permitting multiscale change points,
  heavy tails and dependence}.
\newblock \emph{arXiv preprint arXiv:1910.12486}, 2020.

\bibitem[Cho and Kirch(2021)]{chokirch2021}
Haeran Cho and Claudia Kirch.
\newblock Bootstrap confidence intervals for multiple change points based on
  moving sum procedures.
\newblock \emph{arXiv preprint arXiv:2106.12844}, 2021.

\bibitem[Chu et~al.(1995)Chu, Hornik, and Kaun]{chu1995}
Chia-Shang~J Chu, Kurt Hornik, and Chung-Ming Kaun.
\newblock {MOSUM} tests for parameter constancy.
\newblock \emph{Biometrika}, 82:\penalty0 603--617, 1995.

\bibitem[Chu and Chen(2019)]{chu2019}
Lynna Chu and Hao Chen.
\newblock Asymptotic distribution-free change-point detection for multivariate
  and non-{E}uclidean data.
\newblock \emph{The Annals of Statistics}, 47:\penalty0 382--414, 2019.

\bibitem[Cleynen et~al.(2016)Cleynen, Rigaill, and Koskas]{s3ib}
Alice Cleynen, Guillem Rigaill, and Michel Koskas.
\newblock \emph{{Segmentor3IsBack}: A fast segmentation algorithm}, 2016.
\newblock URL \url{https://CRAN.R-project.org/package=Segmentor3IsBack}.
\newblock {R} package version 2.0.

\bibitem[Cs{\"o}rg{\H{o}} and Horv{\'a}th(1989)]{csorgHo1989invariance}
Mikl{\'o}s Cs{\"o}rg{\H{o}} and Lajos Horv{\'a}th.
\newblock Invariance principles for changepoint problems.
\newblock In \emph{Multivariate statistics and probability}, pages 151--168.
  Elsevier, 1989.

\bibitem[Cs{\"o}rg{\"o} and Horv{\'a}th(1997)]{csorgo1997}
Mikl{\'o}s Cs{\"o}rg{\"o} and Lajos Horv{\'a}th.
\newblock \emph{Limit Theorems in Change-point Analysis}, volume~18.
\newblock John Wiley \& Sons Inc, 1997.

\bibitem[Davis and Yau(2013)]{davis2013}
Richard~A Davis and Chun~Yip Yau.
\newblock Consistency of minimum description length model selection for
  piecewise stationary time series models.
\newblock \emph{Electronic Journal of Statistics}, 7:\penalty0 381--411, 2013.

\bibitem[Davis et~al.(2006)Davis, Lee, and Rodriguez-Yam]{davis2006}
Richard~A Davis, Thomas C~M Lee, and Gabriel~A Rodriguez-Yam.
\newblock Structural break estimation for nonstationary time series models.
\newblock \emph{Journal of the American Statistical Association}, 101:\penalty0
  223--239, 2006.

\bibitem[Davis et~al.(2008)Davis, Lee, and Rodriguez-Yam]{davis2008}
Richard~A Davis, Thomas~CM Lee, and Gabriel~A Rodriguez-Yam.
\newblock Break detection for a class of nonlinear time series models.
\newblock \emph{Journal of Time Series Analysis}, 29:\penalty0 834--867, 2008.

\bibitem[Dehling et~al.(2020)Dehling, Fried, and Wendler]{dehling2020robust}
H~Dehling, R~Fried, and M~Wendler.
\newblock A robust method for shift detection in time series.
\newblock \emph{Biometrika}, 107\penalty0 (3):\penalty0 647--660, 2020.

\bibitem[Dehling et~al.(2013)Dehling, Rooch, and Taqqu]{dehling2013non}
Herold Dehling, Aeneas Rooch, and Murad~S Taqqu.
\newblock Non-parametric change-point tests for long-range dependent data.
\newblock \emph{Scandinavian Journal of Statistics}, 40\penalty0 (1):\penalty0
  153--173, 2013.

\bibitem[Dehling et~al.(2015)Dehling, Fried, Garcia, and
  Wendler]{dehling2015change}
Herold Dehling, Roland Fried, Isabel Garcia, and Martin Wendler.
\newblock Change-point detection under dependence based on two-sample
  {U}-statistics.
\newblock In \emph{Asymptotic Laws and Methods in Stochastics}, pages 195--220.
  Springer, 2015.

\bibitem[Dette et~al.(2020)Dette, Sch{\"u}ler, and Vetter]{dette2018}
Holger Dette, Theresa Sch{\"u}ler, and Mathias Vetter.
\newblock Multiscale change point detection for dependent data.
\newblock \emph{Scandinavian Journal of Statistics}, 47:\penalty0 1243--1274,
  2020.

\bibitem[D\"oring(2010)]{DORING20102003}
Maik D\"oring.
\newblock {Multiple change-point estimation with U-statistics}.
\newblock \emph{Journal of Statistical Planning and Inference}, 140\penalty0
  (7):\penalty0 2003--2017, 2010.

\bibitem[Du et~al.(2016)Du, Kao, and Kou]{du2016}
Chao Du, Chu-Lan~Michael Kao, and SC~Kou.
\newblock Stepwise signal extraction via marginal likelihood.
\newblock \emph{Journal of the American Statistical Association}, 111:\penalty0
  314--330, 2016.

\bibitem[Eckley et~al.(2011)Eckley, Fearnhead, and Killick]{eckley2011}
Idris~A Eckley, Paul Fearnhead, and Rebecca Killick.
\newblock Analysis of changepoint models.
\newblock In David Barber, A.~Taylan Cemgil, and Silvia Chiappa, editors,
  \emph{Bayesian Time Series Models}, pages 205--224. Cambridge University
  Press, 2011.

\bibitem[Efron et~al.(2004)Efron, Hastie, Johnstone, and Tibshirani]{efron2004}
Bradley Efron, Trevor Hastie, Iain Johnstone, and Robert Tibshirani.
\newblock Least angle regression.
\newblock \emph{The Annals of Statistics}, 32:\penalty0 407--499, 2004.

\bibitem[Eichinger and Kirch(2018)]{eichinger2018}
Birte Eichinger and Claudia Kirch.
\newblock {A {MOSUM} procedure for the estimation of multiple random change
  points}.
\newblock \emph{Bernoulli}, 24:\penalty0 526--564, 2018.

\bibitem[Enikeeva and Harchaoui(2019)]{enikeeva2019}
Farida Enikeeva and Zaid Harchaoui.
\newblock High-dimensional change-point detection under sparse alternatives.
\newblock \emph{The Annals of Statistics}, 47:\penalty0 2051--2079, 2019.

\bibitem[Fang et~al.(2020)Fang, Li, and Siegmund]{fang2020}
Xiao Fang, Jian Li, and David Siegmund.
\newblock Segmentation and estimation of change-point models: false positive
  control and confidence regions.
\newblock \emph{The Annals of Statistics}, 48:\penalty0 1615--1647, 2020.

\bibitem[Fearnhead(2006)]{fearnhead2006}
Paul Fearnhead.
\newblock Exact and efficient {Bayesian} inference for multiple changepoint
  problems.
\newblock \emph{Statistics and Computing}, 16:\penalty0 203--213, 2006.

\bibitem[Fearnhead and Rigaill(2019)]{fearnhead2019}
Paul Fearnhead and Guillem Rigaill.
\newblock Changepoint detection in the presence of outliers.
\newblock \emph{Journal of the American Statistical Association}, 114:\penalty0
  169--183, 2019.

\bibitem[Fearnhead and Rigaill(2020)]{fearnhead2020}
Paul Fearnhead and Guillem Rigaill.
\newblock Relating and comparing methods for detecting changes in mean.
\newblock \emph{Stat}, page e291, 2020.

\bibitem[Fearnhead et~al.(2019)Fearnhead, Maidstone, and
  Letchford]{fearnhead2019d}
Paul Fearnhead, Robert Maidstone, and Adam Letchford.
\newblock Detecting changes in slope with an $l_0$ penalty.
\newblock \emph{Journal of Computational and Graphical Statistics},
  28:\penalty0 265--275, 2019.

\bibitem[Ferger(1994)]{ferger1994power}
Dietmar Ferger.
\newblock On the power of nonparametric changepoint-tests.
\newblock \emph{Metrika}, 41\penalty0 (1):\penalty0 277--292, 1994.

\bibitem[Fisch et~al.(2018)Fisch, Eckley, and Fearnhead]{fisch2018}
Alexander Tristan~Maximilian Fisch, Idris~Arthur Eckley, and Paul Fearnhead.
\newblock A linear time method for the detection of point and collective
  anomalies.
\newblock \emph{arXiv preprint arXiv:1806.01947}, 2018.

\bibitem[Frick et~al.(2014)Frick, Munk, and Sieling]{frick2014}
Klaus Frick, Axel Munk, and Hannes Sieling.
\newblock Multiscale change point inference.
\newblock \emph{Journal of the Royal Statistical Society, Series B},
  76:\penalty0 495--580, 2014.

\bibitem[Fromont et~al.(2020)Fromont, Lerasle, and Verzelen]{fromont2020}
M.~Fromont, M.~Lerasle, and N.~Verzelen.
\newblock Optimal change point detection and localization.
\newblock \emph{arXiv preprint arXiv:2010.11470}, 2020.

\bibitem[Fryzlewicz(2014)]{fryzlewicz2014}
Piotr Fryzlewicz.
\newblock Wild binary segmentation for multiple change-point detection.
\newblock \emph{The Annals of Statistics}, 42:\penalty0 2243--2281, 2014.

\bibitem[Fryzlewicz(2018)]{fryzlewicz2017}
Piotr Fryzlewicz.
\newblock Tail-greedy bottom-up data decompositions and fast multiple
  change-point detection.
\newblock \emph{The Annals of Statistics}, pages 3390--3421, 2018.

\bibitem[Fryzlewicz(2020{\natexlab{a}})]{fryzlewicz2020}
Piotr Fryzlewicz.
\newblock Detecting possibly frequent change-points: {Wild Binary Segmentation}
  2 and steepest-drop model selection.
\newblock \emph{Journal of the Korean Statistical Society}, 49:\penalty0 1--44,
  2020{\natexlab{a}}.

\bibitem[Fryzlewicz(2020{\natexlab{b}})]{fryzlewicz2020narrowest}
Piotr Fryzlewicz.
\newblock Narrowest significance pursuit: inference for multiple change-points
  in linear models.
\newblock \emph{arXiv preprint arXiv:2009.05431}, 2020{\natexlab{b}}.

\bibitem[Fryzlewicz and {Subba Rao}(2014)]{fryzlewicz2014b}
Piotr Fryzlewicz and Suhasini {Subba Rao}.
\newblock Multiple-change-point detection for auto-regressive conditional
  heteroscedastic processes.
\newblock \emph{Journal of the Royal Statistical Society, Series B},
  76:\penalty0 903--924, 2014.

\bibitem[Fuh(2004)]{fuh2004}
Cheng-Der Fuh.
\newblock Asymptotic operating characteristics of an optimal change point
  detection in hidden {Markov} models.
\newblock \emph{The Annals of Statistics}, 32:\penalty0 2305--2339, 2004.

\bibitem[Garreau and Arlot(2018)]{garreau2018}
Damien Garreau and Sylvain Arlot.
\newblock Consistent change-point detection with kernels.
\newblock \emph{Electronic Journal of Statistics}, 12:\penalty0 4440--4486,
  2018.

\bibitem[Gombay(2001)]{gombay2001u}
Edit Gombay.
\newblock {U-statistics for change under alternatives}.
\newblock \emph{Journal of Multivariate Analysis}, 78\penalty0 (1):\penalty0
  139--158, 2001.

\bibitem[G{\'o}recki et~al.(2018)G{\'o}recki, Horv{\'a}th, and
  Kokoszka]{gorecki2018change}
Tomasz G{\'o}recki, Lajos Horv{\'a}th, and Piotr Kokoszka.
\newblock Change point detection in heteroscedastic time series.
\newblock \emph{Econometrics and Statistics}, 7:\penalty0 63--88, 2018.

\bibitem[Grunwald(2004)]{grunwald2004}
Peter Grunwald.
\newblock A tutorial introduction to the minimum description length principle.
\newblock \emph{arXiv preprint arXiv:math/0406077}, 2004.

\bibitem[Harchaoui and L{\'e}vy-Leduc(2010)]{harchaoui2010}
Za{\i}d Harchaoui and C{\'e}line L{\'e}vy-Leduc.
\newblock Multiple change-point estimation with a total variation penalty.
\newblock \emph{Journal of the American Statistical Association}, 105:\penalty0
  1480--1493, 2010.

\bibitem[Haynes et~al.(2017{\natexlab{a}})Haynes, Eckley, and
  Fearnhead]{haynes2017a}
Kaylea Haynes, Idris~A Eckley, and Paul Fearnhead.
\newblock Computationally efficient changepoint detection for a range of
  penalties.
\newblock \emph{Journal of Computational and Graphical Statistics},
  26:\penalty0 134--143, 2017{\natexlab{a}}.

\bibitem[Haynes et~al.(2017{\natexlab{b}})Haynes, Fearnhead, and
  Eckley]{haynes2017b}
Kaylea Haynes, Paul Fearnhead, and Idris~A Eckley.
\newblock A computationally efficient nonparametric approach for changepoint
  detection.
\newblock \emph{Statistics and Computing}, 27:\penalty0 1293--1305,
  2017{\natexlab{b}}.

\bibitem[Heunis(2003)]{Heunis}
A.~J. Heunis.
\newblock Strong invariance principle for singular diffusions.
\newblock \emph{Stochastic Processes and their Applications}, 104:\penalty0
  57--80, 2003.

\bibitem[Horv{\'a}th and Hu{\v{s}}kov{\'a}(2012)]{horvath2012}
L.~Horv{\'a}th and M.~Hu{\v{s}}kov{\'a}.
\newblock Change-point detection in panel data.
\newblock \emph{Journal of Time Series Analysis}, 33:\penalty0 631--648, 2012.

\bibitem[Horv{\'a}th and Rice(2014)]{horvath2014}
Lajos Horv{\'a}th and Gregory Rice.
\newblock Extensions of some classical methods in change point analysis.
\newblock \emph{TEST}, 23:\penalty0 1--37, 2014.

\bibitem[Horv{\'a}th and Steinebach(2000)]{horvath2000testing}
Lajos Horv{\'a}th and Josef Steinebach.
\newblock Testing for changes in the mean or variance of a stochastic process
  under weak invariance.
\newblock \emph{Journal of Statistical Planning and Inference}, 91:\penalty0
  365--376, 2000.

\bibitem[Hu{\v{s}}kov{\'a}(1990{\natexlab{a}})]{huvskova1990asymptotics}
Marie Hu{\v{s}}kov{\'a}.
\newblock Asymptotics for robust {MOSUM}.
\newblock \emph{Commentationes Mathematicae Universitatis Carolinae},
  31\penalty0 (2):\penalty0 345--356, 1990{\natexlab{a}}.

\bibitem[Hu{\v{s}}kov{\'a}(1990{\natexlab{b}})]{huvskova1990some}
Marie Hu{\v{s}}kov{\'a}.
\newblock Some asymptotic results for robust procedures for testing the
  constancy of regression models over time.
\newblock \emph{Kybernetika}, 26\penalty0 (5):\penalty0 392--403,
  1990{\natexlab{b}}.

\bibitem[Hu{\v{s}}kov{\'a}(1996)]{huvskova1996tests}
Marie Hu{\v{s}}kov{\'a}.
\newblock Tests and estimators for the change point problem based on
  {M}-statistics.
\newblock \emph{Statistics \& Risk Modeling}, 14:\penalty0 115--136, 1996.

\bibitem[Hu{\v{s}}kov{\'a}(2004)]{huvskova2004}
Marie Hu{\v{s}}kov{\'a}.
\newblock Permutation principle and bootstrap in change point analysis.
\newblock In L.~Horv{\'a}th and B.~Szyszkowicz, editors, \emph{Asymptotic
  Methods in Stochastics. Festschrift for Mikl{\'o}s Cs{\"o}rg{\"o}},
  volume~44, pages 273--292. American Mathematical Society, 2004.

\bibitem[Hu{\v{s}}kov{\'a}(2013)]{huvskova2013robust}
Marie Hu{\v{s}}kov{\'a}.
\newblock Robust change point analysis.
\newblock In \emph{Robustness and Complex Data Structures}, pages 171--190.
  Springer, 2013.

\bibitem[Hu{\v{s}}kov{\'a} and Kirch(2008)]{huvskova2008}
Marie Hu{\v{s}}kov{\'a} and Claudia Kirch.
\newblock Bootstrapping confidence intervals for the change-point of time
  series.
\newblock \emph{Journal of Time Series Analysis}, 29:\penalty0 947--972, 2008.

\bibitem[Hu{\v{s}}kov{\'a} and Kirch(2010)]{huvskova2010}
Marie Hu{\v{s}}kov{\'a} and Claudia Kirch.
\newblock A note on studentized confidence intervals for the change-point.
\newblock \emph{Computational Statistics}, 25:\penalty0 269--289, 2010.

\bibitem[Hu{\v{s}}kov{\'a} and Maru{\v{s}}iakov{\'a}(2012)]{huvskova2012m}
Marie Hu{\v{s}}kov{\'a} and Miriam Maru{\v{s}}iakov{\'a}.
\newblock {M-procedures for detection of changes for dependent observations}.
\newblock \emph{Communications in Statistics -- Simulation and Computation},
  41\penalty0 (7):\penalty0 1032--1050, 2012.

\bibitem[Hu{\v{s}}kov{\'a} and
  Meintanis(2006{\natexlab{a}})]{huvskova2006change}
Marie Hu{\v{s}}kov{\'a} and Simos~G Meintanis.
\newblock Change point analysis based on empirical characteristic functions.
\newblock \emph{Metrika}, 63:\penalty0 145--168, 2006{\natexlab{a}}.

\bibitem[Hu{\v{s}}kov{\'a} and
  Meintanis(2006{\natexlab{b}})]{huvskova2006change1}
Marie Hu{\v{s}}kov{\'a} and Simos~G Meintanis.
\newblock Change-point analysis based on empirical characteristic functions of
  ranks.
\newblock \emph{Sequential Analysis}, 25\penalty0 (4):\penalty0 421--436,
  2006{\natexlab{b}}.

\bibitem[Hu{\v{s}}kov{\'a} and Picek(2002)]{huvskova2002m}
Marie Hu{\v{s}}kov{\'a} and Jan Picek.
\newblock {M-tests for detection of structural changes in regression}.
\newblock In \emph{Statistical data analysis based on the L1-norm and related
  methods}, pages 213--227. Springer, 2002.

\bibitem[Hu{\v{s}}kov{\'a} and Slab{\`y}(2001)]{huvskova2001}
Marie Hu{\v{s}}kov{\'a} and Ale{\v{s}} Slab{\`y}.
\newblock Permutation tests for multiple changes.
\newblock \emph{Kybernetika}, 37:\penalty0 605--622, 2001.

\bibitem[Hu{\v{s}}kov{\'a} et~al.(2007)Hu{\v{s}}kov{\'a},
  Pr{\'a}{\v{s}}kov{\'a}, and Steinebach]{huvskova2007detection}
Marie Hu{\v{s}}kov{\'a}, Zuzana Pr{\'a}{\v{s}}kov{\'a}, and Josef Steinebach.
\newblock On the detection of changes in autoregressive time series {I.
  Asymptotics}.
\newblock \emph{Journal of Statistical Planning and Inference}, 137:\penalty0
  1243--1259, 2007.

\bibitem[Hyun et~al.(2021)Hyun, Lin, G'Sell, and Tibshirani]{hyun2018post}
Sangwon Hyun, Kevin~Z Lin, Max G'Sell, and Ryan~J Tibshirani.
\newblock Post-selection inference for changepoint detection algorithms with
  application to copy number variation data.
\newblock \emph{Biometrics}, 2021.

\bibitem[Jackson et~al.(2005)Jackson, Scargle, Barnes, Arabhi, Alt, Gioumousis,
  Gwin, Sangtrakulcharoen, Tan, and Tsai]{jackson2005}
Brad Jackson, Jeffrey~D Scargle, David Barnes, Sundararajan Arabhi, Alina Alt,
  Peter Gioumousis, Elyus Gwin, Paungkaew Sangtrakulcharoen, Linda Tan, and
  Tun~Tao Tsai.
\newblock An algorithm for optimal partitioning of data on an interval.
\newblock \emph{IEEE Signal Processing Letters}, 12:\penalty0 105--108, 2005.

\bibitem[Jandhyala et~al.(2013)Jandhyala, Fotopoulos, MacNeill, and
  Liu]{jandhyala2013inference}
Venkata Jandhyala, Stergios Fotopoulos, Ian MacNeill, and Pengyu Liu.
\newblock Inference for single and multiple change-points in time series.
\newblock \emph{Journal of Time Series Analysis}, 34\penalty0 (4):\penalty0
  423--446, 2013.

\bibitem[Jewell et~al.(2019)Jewell, Fearnhead, and Witten]{jewell2019testing}
Sean Jewell, Paul Fearnhead, and Daniela Witten.
\newblock Testing for a change in mean after changepoint detection.
\newblock \emph{arXiv preprint arXiv:1910.04291}, 2019.

\bibitem[Jirak(2015)]{jirak2014}
Moritz Jirak.
\newblock Uniform change point tests in high dimension.
\newblock \emph{The Annals of Statistics}, 43:\penalty0 2451--2483, 2015.

\bibitem[Kaul et~al.(2019)Kaul, Jandhyala, and Fotopoulos]{kaul2019}
Abhishek Kaul, Venkata~K Jandhyala, and Stergios~B Fotopoulos.
\newblock Detection and estimation of parameters in high dimensional multiple
  change point regression models via $\ell_1/\ell_0$ regularization and
  discrete optimization.
\newblock \emph{arXiv preprint arXiv:1906.04396}, 2019.

\bibitem[Killick et~al.(2012)Killick, Fearnhead, and Eckley]{killick2012}
Rebecca Killick, Paul Fearnhead, and Idris~A Eckley.
\newblock Optimal detection of changepoints with a linear computational cost.
\newblock \emph{Journal of the American Statistical Association}, 107:\penalty0
  1590--1598, 2012.

\bibitem[Killick et~al.(2016)Killick, Haynes, and Eckley]{changepoint}
Rebecca Killick, Kaylea Haynes, and Idris~A. Eckley.
\newblock \emph{{changepoint}: An {R} package for changepoint analysis}, 2016.
\newblock URL \url{https://CRAN.R-project.org/package=changepoint}.
\newblock {R} package version 2.2.2.

\bibitem[Kirch and Kamgaing(2014)]{kirch2014detection}
Claudia Kirch and J~Tadjuidje Kamgaing.
\newblock Detection of change points in discrete valued time series.
\newblock In Richard~A. Davis, Scott~H. Holan, Robert Lund, and Nalini
  Ravishanker, editors, \emph{Handbook of Discrete Valued Time Series}. Chapman
  and Hall/CRC, New York, 2014.

\bibitem[Kirch and Kamgaing(2012)]{kirch2012testing}
Claudia Kirch and Joseph~Tadjuidje Kamgaing.
\newblock Testing for parameter stability in nonlinear autoregressive models.
\newblock \emph{Journal of Time Series Analysis}, 33:\penalty0 365--385, 2012.

\bibitem[Kirch and Klein(2021)]{klein2020}
Claudia Kirch and Philipp Klein.
\newblock Moving sum data segmentation for stochastics processes based on
  invariance.
\newblock \emph{arXiv preprint arXiv:2101.04651}, 2021.

\bibitem[Kirch and Reckr{\"u}hm(2021)]{kirchreckruehm2020}
Claudia Kirch and Kerstin Reckr{\"u}hm.
\newblock Data segmentation for time series based on a general moving sum
  approach.
\newblock In preparation, 2021.

\bibitem[Kirch et~al.(2015)Kirch, Muhsal, and Ombao]{kirch2015eeg}
Claudia Kirch, Birte Muhsal, and Hernando Ombao.
\newblock Detection of changes in multivariate time series with application to
  {EEG} data.
\newblock \emph{Journal of the American Statistical Association}, 110:\penalty0
  1197--1216, 2015.

\bibitem[Koml{\'o}s et~al.(1975)Koml{\'o}s, Major, and Tusn{\'a}dy]{komlos1975}
J{\'a}nos Koml{\'o}s, P{\'e}ter Major, and G{\'a}bor Tusn{\'a}dy.
\newblock {An approximation of partial sums of independent RV's, and the sample
  DF. I}.
\newblock \emph{Zeitschrift f{\"u}r Wahrscheinlichkeitstheorie und verwandte
  Gebiete}, 32:\penalty0 111--131, 1975.

\bibitem[Koml{\'o}s et~al.(1976)Koml{\'o}s, Major, and Tusn{\'a}dy]{komlos1976}
J{\'a}nos Koml{\'o}s, P{\'e}ter Major, and G{\'a}bor Tusn{\'a}dy.
\newblock {An approximation of partial sums of independent RV's, and the sample
  DF. II}.
\newblock \emph{Zeitschrift f{\"u}r Wahrscheinlichkeitstheorie und verwandte
  Gebiete}, 34:\penalty0 33--58, 1976.

\bibitem[Korostelev(1987)]{korostelev1987}
A.~Korostelev.
\newblock On minimax estimation of a discontinuous signal.
\newblock \emph{Theory of Probability and its Applications}, 32:\penalty0
  727--730, 1987.

\bibitem[Kov{\'a}cs et~al.(2020)Kov{\'a}cs, Li, B{\"u}hlmann, and
  Munk]{kovacs2020}
Solt Kov{\'a}cs, Housen Li, Peter B{\"u}hlmann, and Axel Munk.
\newblock Seeded binary segmentation: A general methodology for fast and
  optimal change point detection.
\newblock \emph{arXiv preprint arXiv:2002.06633}, 2020.

\bibitem[Kuelbs and Philipp(1980)]{kuelbs1980}
James Kuelbs and Walter Philipp.
\newblock {Almost sure invariance principles for partial sums of mixing
  $B$-valued random variables}.
\newblock \emph{The Annals of Probability}, 8:\penalty0 1003--1036, 1980.

\bibitem[K{\"u}hn(2001)]{kuhn2001estimator}
Christoph K{\"u}hn.
\newblock An estimator of the number of change points based on a weak
  invariance principle.
\newblock \emph{Statistics \& Probability Letters}, 51:\penalty0 189--196,
  2001.

\bibitem[Lavielle and Moulines(2000)]{lavielle2000}
Marc Lavielle and Eric Moulines.
\newblock Least-squares estimation of an unknown number of shifts in a time
  series.
\newblock \emph{Journal of Time Series Analysis}, 21:\penalty0 33--59, 2000.

\bibitem[Lebarbier(2005)]{lebarbier2005detecting}
{\'E}milie Lebarbier.
\newblock Detecting multiple change-points in the mean of {G}aussian process by
  model selection.
\newblock \emph{Signal Processing}, 85\penalty0 (4):\penalty0 717--736, 2005.

\bibitem[Lee(1995)]{lee1995}
Chung-Bow Lee.
\newblock Estimating the number of change points in a sequence of independent
  normal random variables.
\newblock \emph{Statistics \& Probability Letters}, 25:\penalty0 241--248,
  1995.

\bibitem[Leonardi and B{\"u}hlmann(2016)]{leonardi2016}
Florencia Leonardi and Peter B{\"u}hlmann.
\newblock Computationally efficient change point detection for high-dimensional
  regression.
\newblock \emph{arXiv preprint arXiv:1601.03704}, 2016.

\bibitem[Li and Sieling(2017)]{fdrseg}
Housen Li and Hannes Sieling.
\newblock \emph{{FDRSeg}: {FDR}-control in multiscale change-point
  segmentation}, 2017.
\newblock URL \url{https://CRAN.R-project.org/package=FDRSeg}.
\newblock {R} package version 1.0-3.

\bibitem[Li et~al.(2016)Li, Munk, and Sieling]{li2016}
Housen Li, Axel Munk, and Hannes Sieling.
\newblock {FDR-control in multiscale change-point segmentation}.
\newblock \emph{Electronic Journal of Statistics}, 10:\penalty0 918--959, 2016.

\bibitem[Li et~al.(2019)Li, Guo, and Munk]{li2019}
Housen Li, Qinghai Guo, and Axel Munk.
\newblock Multiscale change-point segmentation: Beyond step functions.
\newblock \emph{Electronic Journal of Statistics}, 13:\penalty0 3254--3296,
  2019.

\bibitem[Lin et~al.(2017)Lin, Sharpnack, Rinaldo, and Tibshirani]{lin2017}
Kevin Lin, James~L Sharpnack, Alessandro Rinaldo, and Ryan~J Tibshirani.
\newblock A sharp error analysis for the fused {Lasso}, with application to
  approximate changepoint screening.
\newblock In \emph{Advances in Neural Information Processing Systems}, pages
  6884--6893, 2017.

\bibitem[Liu et~al.(2021)Liu, Gao, and Samworth]{liu2019}
Haoyang Liu, Chao Gao, and Richard~J Samworth.
\newblock Minimax rates in sparse, high-dimensional change point detection.
\newblock \emph{The Annals of Statistics}, 49\penalty0 (2):\penalty0
  1081--1112, 2021.

\bibitem[Liu and Chen(2020)]{liu2020fast}
Yi-Wei Liu and Hao Chen.
\newblock A fast and efficient change-point detection framework for modern
  data.
\newblock \emph{arXiv preprint arXiv:2006.13450}, 2020.

\bibitem[Lu et~al.(2020)Lu, Banerjee, and Michailidis]{lu2017}
Zhiyuan Lu, Moulinath Banerjee, and George Michailidis.
\newblock Intelligent sampling for multiple change-points in exceedingly long
  time series with rate guarantees.
\newblock \emph{arXiv preprint arXiv:1710.07420}, 2020.

\bibitem[Maeng and Fryzlewicz(2019)]{maeng2019}
Hyeyoung Maeng and Piotr Fryzlewicz.
\newblock Detecting linear trend changes and point anomalies in data sequences.
\newblock \emph{arXiv preprint arXiv:1906.01939}, 2019.

\bibitem[Maidstone et~al.(2017)Maidstone, Hocking, Rigaill, and
  Fearnhead]{maidstone2017}
Robert Maidstone, Toby Hocking, Guillem Rigaill, and Paul Fearnhead.
\newblock On optimal multiple changepoint algorithms for large data.
\newblock \emph{Statistics and Computing}, 27:\penalty0 519--533, 2017.

\bibitem[Maru{\v{s}}iakov{\'a}(2009)]{maruvsiakova2009tests}
Miriam Maru{\v{s}}iakov{\'a}.
\newblock \emph{Tests for multiple changes in linear regression models}.
\newblock PhD thesis, Charles University in Prague, 2009.

\bibitem[Matteson and James(2014)]{matteson2014}
David~S. Matteson and Nicholas~A. James.
\newblock A nonparametric approach for multiple change point analysis of
  multivariate data.
\newblock \emph{Journal of the American Statistical Association}, 109:\penalty0
  334--345, 2014.

\bibitem[Meier et~al.(2021{\natexlab{a}})Meier, Cho, and Kirch]{mosum}
Alexander Meier, Haeran Cho, and Claudia Kirch.
\newblock \emph{mosum: Moving sum based procedures for changes in the mean},
  2021{\natexlab{a}}.
\newblock URL \url{https://CRAN.R-project.org/package=mosum}.
\newblock {R} package version 1.2.6.

\bibitem[Meier et~al.(2021{\natexlab{b}})Meier, Kirch, and Cho]{meier2021mosum}
Alexander Meier, Claudia Kirch, and Haeran Cho.
\newblock mosum: A package for moving sums in change-point analysis.
\newblock \emph{Journal of Statistical Software}, 97\penalty0 (1):\penalty0
  1--42, 2021{\natexlab{b}}.

\bibitem[Messer et~al.(2014)Messer, Kirchner, Schiemann, Roeper, Neininger, and
  Schneider]{messer2014}
Michael Messer, Marietta Kirchner, Julia Schiemann, Jochen Roeper, Ralph
  Neininger, and Gaby Schneider.
\newblock A multiple filter test for the detection of rate changes in renewal
  processes with varying variance.
\newblock \emph{The Annals of Applied Statistics}, 8:\penalty0 2027--2067,
  2014.

\bibitem[Messer et~al.(2018)Messer, Albert, and Schneider]{messer2018}
Michael Messer, Stefan Albert, and Gaby Schneider.
\newblock The multiple filter test for change point detection in time series.
\newblock \emph{Metrika}, 81:\penalty0 589--607, 2018.

\bibitem[Mohr and Neumeyer(2020)]{mohr2020consistent}
Maria Mohr and Natalie Neumeyer.
\newblock Consistent nonparametric change point detection combining {CUSUM} and
  marked empirical processes.
\newblock \emph{Electronic Journal of Statistics}, 14:\penalty0 2238--2271,
  2020.

\bibitem[Ninomiya(2005)]{ninomiya2005}
Yoshiyuki Ninomiya.
\newblock Information criterion for {Gaussian} change-point model.
\newblock \emph{Statistics \& Probability Letters}, 72:\penalty0 237--247,
  2005.

\bibitem[Niu and Zhang(2012)]{niu2012}
Yue~S Niu and Heping Zhang.
\newblock The screening and ranking algorithm to detect {DNA} copy number
  variations.
\newblock \emph{The Annals of Applied Statistics}, 6:\penalty0 1306--1326,
  2012.

\bibitem[Olshen et~al.(2004)Olshen, Venkatraman, Lucito, and
  Wigler]{olshen2004}
Adam~B Olshen, ES~Venkatraman, Robert Lucito, and Michael Wigler.
\newblock Circular binary segmentation for the analysis of array-based {DNA}
  copy number data.
\newblock \emph{Biostatistics}, 5:\penalty0 557--572, 2004.

\bibitem[Orasch(2004)]{orasch2004using}
M~Orasch.
\newblock {Using U-statistcs based processes to detect multiple change-points}.
\newblock In L.~Horv{\'a}th and B.~Szyszkowicz, editors, \emph{Asymptotic
  Methods in Stochastics. Festschrift for Mikl{\'o}s Cs{\"o}rg{\"o}},
  volume~44, pages 315--334. American Mathematical Society, 2004.

\bibitem[Padilla et~al.(2019)Padilla, Yu, Wang, and Rinaldo]{padilla2019}
Oscar Hernan~Madrid Padilla, Yi~Yu, Daren Wang, and Alessandro Rinaldo.
\newblock Optimal nonparametric change point detection and localization.
\newblock \emph{arXiv preprint arXiv:1905.10019}, 2019.

\bibitem[Page(1954)]{page1954}
Ewan~S Page.
\newblock Continuous inspection schemes.
\newblock \emph{Biometrika}, 41:\penalty0 100--115, 1954.

\bibitem[Pan and Chen(2006)]{pan2006}
Jianmin Pan and Jiahua Chen.
\newblock Application of modified information criterion to multiple change
  point problems.
\newblock \emph{Journal of Multivariate Analysis}, 97:\penalty0 2221--2241,
  2006.

\bibitem[Pein et~al.(2017)Pein, Sieling, and Munk]{pein2015}
Florian Pein, Hannes Sieling, and Axel Munk.
\newblock Heterogeneous change point inference.
\newblock \emph{Journal of the Royal Statistical Society, Series B},
  79:\penalty0 1207--1227, 2017.

\bibitem[Pein et~al.(2019)Pein, Hotz, Sieling, and Aspelmeier]{stepR}
Florian Pein, Thomas Hotz, Hannes Sieling, and Timo Aspelmeier.
\newblock \emph{{stepR}: Multiscale change-point inference}, 2019.
\newblock URL \url{https://CRAN.R-project.org/package=stepR}.
\newblock {R} package version 2.0-3.

\bibitem[Perron(2006)]{perron2006}
Pierre Perron.
\newblock Dealing with structural breaks.
\newblock \emph{Palgrave Handbook of Econometrics}, 1:\penalty0 278--352, 2006.

\bibitem[Pr{\'a}{\v{s}}kov{\'a} and Chochola(2014)]{pravskova2014m}
Zuzana Pr{\'a}{\v{s}}kov{\'a} and Ond{\v{r}}ej Chochola.
\newblock {M-procedures for detection of a change under weak dependence}.
\newblock \emph{Journal of Statistical Planning and Inference}, 149:\penalty0
  60--76, 2014.

\bibitem[Preuss et~al.(2015)Preuss, Puchstein, and Dette]{preuss2015}
Philip Preuss, Ruprecht Puchstein, and Holger Dette.
\newblock Detection of multiple structural breaks in multivariate time series.
\newblock \emph{Journal of the American Statistical Association}, 110:\penalty0
  654--668, 2015.

\bibitem[Reckr{\"u}hm(2019)]{reckruhm2019}
Kerstin Reckr{\"u}hm.
\newblock \emph{Estimating multiple structural breaks in time series -- a
  generalized {MOSUM} approach based on estimating functions}.
\newblock PhD thesis, Otto-von-Guericke University, Magdeburg, Germany, 2019.

\bibitem[Reeves et~al.(2007)Reeves, Chen, Wang, Lund, and Lu]{reeves2007}
Jaxk Reeves, Jien Chen, Xiaolan~L Wang, Robert Lund, and Qi~Qi Lu.
\newblock A review and comparison of changepoint detection techniques for
  climate data.
\newblock \emph{Journal of Applied Meteorology and Climatology}, 46:\penalty0
  900--915, 2007.

\bibitem[Rigaill(2015)]{rigaill2010}
Guillem Rigaill.
\newblock A pruned dynamic programming algorithm to recover the best
  segmentations with 1 to {K}\_max change-points.
\newblock \emph{Journal de la Soci{\'e}t{\'e} Fran{\c{c}}aise de Statistique},
  156:\penalty0 180--205, 2015.

\bibitem[Rigaill and Hocking(2019)]{fpop}
Guillem Rigaill and Toby~Dylan Hocking.
\newblock \emph{fpop: Segmentation using Optimal Partitioning and Function
  Pruning}, 2019.
\newblock URL \url{https://R-Forge.R-project.org/projects/opfp/}.
\newblock {R} package version 2019.01.22/r56.

\bibitem[Rigaill et~al.(2012)Rigaill, Lebarbier, and Robin]{rigaill2012}
Guillem Rigaill, {\'E}milie Lebarbier, and Stephane Robin.
\newblock Exact posterior distributions and model selection criteria for
  multiple change-point detection problems.
\newblock \emph{Statistics and Computing}, 22:\penalty0 917--929, 2012.

\bibitem[Rinaldo(2009)]{rinaldo2009}
Alessandro Rinaldo.
\newblock Properties and refinements of the fused {Lasso}.
\newblock \emph{The Annals of Statistics}, 37:\penalty0 2922--2952, 2009.

\bibitem[Rissanen(1978)]{rissanen1978}
Jorma Rissanen.
\newblock Modeling by shortest data description.
\newblock \emph{Automatica}, 14:\penalty0 465--471, 1978.

\bibitem[Romano et~al.(2021)Romano, Rigaill, Runge, and
  Fearnhead]{romano2021detecting}
Gaetano Romano, Guillem Rigaill, Vincent Runge, and Paul Fearnhead.
\newblock Detecting abrupt changes in the presence of local fluctuations and
  autocorrelated noise.
\newblock \emph{Journal of the American Statistical Association (to appear)},
  2021.

\bibitem[Safikhani and Shojaie(2020)]{safikhani2020}
Abolfazl Safikhani and Ali Shojaie.
\newblock Joint structural break detection and parameter estimation in
  high-dimensional nonstationary {VAR} models.
\newblock \emph{Journal of the American Statistical Association}, pages 1--14,
  2020.

\bibitem[Schwarz(1978)]{schwarz1978}
Gideon Schwarz.
\newblock Estimating the dimension of a model.
\newblock \emph{The Annals of Statistics}, 6:\penalty0 461--464, 1978.

\bibitem[Scott and Knott(1974)]{scott1974}
Andrew~Jhon Scott and M~Knott.
\newblock A cluster analysis method for grouping means in the analysis of
  variance.
\newblock \emph{Biometrics}, 6:\penalty0 507--512, 1974.

\bibitem[Sharipov et~al.(2016)Sharipov, Tewes, and
  Wendler]{sharipov2016sequential}
Olimjon Sharipov, Johannes Tewes, and Martin Wendler.
\newblock Sequential block bootstrap in a {Hilbert} space with application to
  change point analysis.
\newblock \emph{Canadian Journal of Statistics}, 44:\penalty0 300--322, 2016.

\bibitem[Shin et~al.(2020)Shin, Wu, and Hao]{shin2020}
Seung~Jun Shin, Yichao Wu, and Ning Hao.
\newblock A backward procedure for change-point detection with applications to
  copy number variation detection.
\newblock \emph{Canadian Journal of Statistics}, 48:\penalty0 366--385, 2020.

\bibitem[Steinebach and Eastwood(1996)]{SteinebachEastwood}
J.~Steinebach and V.~R. Eastwood.
\newblock Extreme value asymptotics for multivariate renewal processes.
\newblock \emph{Journal of Multivariate Analysis}, 56:\penalty0 284--302, 1996.

\bibitem[Steland(2020)]{steland2020testing}
Ansgar Steland.
\newblock Testing and estimating change-points in the covariance matrix of a
  high-dimensional time series.
\newblock \emph{Journal of Multivariate Analysis}, 177:\penalty0 104582, 2020.

\bibitem[Stoehr et~al.(2021)Stoehr, Aston, and Kirch]{stoehr2020detecting}
Christina Stoehr, John~AD Aston, and Claudia Kirch.
\newblock Detecting changes in the covariance structure of functional time
  series with application to fmri data.
\newblock \emph{Econometrics and Statistics}, 18:\penalty0 44--62, 2021.

\bibitem[Tartakovsky(2019)]{tartakovsky2019sequential}
Alexander Tartakovsky.
\newblock \emph{Sequential Change Detection and Hypothesis Testing: General
  Non-iid Stochastic Models and Asymptotically Optimal Rules}.
\newblock CRC Press, 2019.

\bibitem[Tartakovsky et~al.(2014)Tartakovsky, Nikiforov, and
  Basseville]{tartakovsky2014sequential}
Alexander Tartakovsky, Igor Nikiforov, and Michele Basseville.
\newblock \emph{Sequential analysis: Hypothesis testing and changepoint
  detection}.
\newblock CRC Press, 2014.

\bibitem[Tecuapetla-G{\'o}mez and Munk(2017)]{tecuapetla2017}
Inder Tecuapetla-G{\'o}mez and Axel Munk.
\newblock Autocovariance estimation in regression with a discontinuous signal
  and m-dependent errors: A difference-based approach.
\newblock \emph{Scandinavian Journal of Statistics}, 44:\penalty0 346--368,
  2017.

\bibitem[Tibshirani et~al.(2005)Tibshirani, Saunders, Rosset, Zhu, and
  Knight]{tibshirani2005}
Robert Tibshirani, Michael Saunders, Saharon Rosset, Ji~Zhu, and Keith Knight.
\newblock Sparsity and smoothness via the fused {Lasso}.
\newblock \emph{Journal of the Royal Statistical Society, Series B},
  67:\penalty0 91--108, 2005.

\bibitem[Tibshirani(2014)]{tibshirani2014}
Ryan~J Tibshirani.
\newblock Adaptive piecewise polynomial estimation via trend filtering.
\newblock \emph{The Annals of Statistics}, 42:\penalty0 285--323, 2014.

\bibitem[Tickle et~al.(2020)Tickle, Eckley, Fearnhead, and Haynes]{tickle2020}
SO~Tickle, IA~Eckley, Paul Fearnhead, and Kaylea Haynes.
\newblock Parallelization of a common changepoint detection method.
\newblock \emph{Journal of Computational and Graphical Statistics},
  29:\penalty0 149--161, 2020.

\bibitem[Titsias et~al.(2016)Titsias, Holmes, and Yau]{titsias2016}
Michalis~K Titsias, Christopher~C Holmes, and Christopher Yau.
\newblock Statistical inference in hidden {Markov} models using k-segment
  constraints.
\newblock \emph{Journal of the American Statistical Association}, 111:\penalty0
  200--215, 2016.

\bibitem[Truong et~al.(2020)Truong, Oudre, and Vayatis]{truong2020}
Charles Truong, Laurent Oudre, and Nicolas Vayatis.
\newblock Selective review of offline change point detection methods.
\newblock \emph{Signal Processing}, 167:\penalty0 107299, 2020.

\bibitem[Venkatraman(1992)]{venkatraman1992}
E.~S. Venkatraman.
\newblock \emph{Consistency Results in Multiple Change-point Problems}.
\newblock PhD thesis, Stanford University, 1992.

\bibitem[Venkatraman and Olshen(2007)]{venkatraman2007}
ES~Venkatraman and Adam~B Olshen.
\newblock A faster circular binary segmentation algorithm for the analysis of
  array {CGH} data.
\newblock \emph{Bioinformatics}, 23:\penalty0 657--663, 2007.

\bibitem[Vogel and Wendler(2017)]{vogel2017studentized}
Daniel Vogel and Martin Wendler.
\newblock {Studentized U-quantile processes under dependence with applications
  to change-point analysis}.
\newblock \emph{Bernoulli}, 23\penalty0 (4B):\penalty0 3114--3144, 2017.

\bibitem[Vostrikova(1981)]{vostrikova1981}
L.~J. Vostrikova.
\newblock Detecting `disorder' in multidimensional random processes.
\newblock \emph{Soviet Doklady Mathematics}, 24:\penalty0 55--59, 1981.

\bibitem[Wang et~al.(2019{\natexlab{a}})Wang, Lin, and Willett]{wang2019b}
Daren Wang, Kevin Lin, and Rebecca Willett.
\newblock Statistically and computationally efficient change point localization
  in regression settings.
\newblock \emph{arXiv preprint arXiv:1906.11364}, 2019{\natexlab{a}}.

\bibitem[Wang et~al.(2019{\natexlab{b}})Wang, Yu, Rinaldo, and
  Willett]{wang2019}
Daren Wang, Yi~Yu, Alessandro Rinaldo, and Rebecca Willett.
\newblock Localizing changes in high-dimensional vector autoregressive
  processes.
\newblock \emph{arXiv preprint arXiv:1909.06359}, 2019{\natexlab{b}}.

\bibitem[Wang et~al.(2020)Wang, Yu, and Rinaldo]{wang2018d}
Daren Wang, Yi~Yu, and Alessandro Rinaldo.
\newblock Univariate mean change point detection: Penalization, cusum and
  optimality.
\newblock \emph{Electronic Journal of Statistics}, 14:\penalty0 1917--1961,
  2020.

\bibitem[Wang et~al.(2021)Wang, Yu, and Rinaldo]{wang2017}
Daren Wang, Yi~Yu, and Alessandro Rinaldo.
\newblock Optimal covariance change point localization in high dimension.
\newblock \emph{Bernoulli}, 27:\penalty0 554--575, 2021.

\bibitem[Wang et~al.(2019{\natexlab{c}})Wang, Volgushev, and Shao]{wvs2019}
Runmin Wang, Stanislav Volgushev, and Xiaofeng Shao.
\newblock Inference for change points in high dimensional data.
\newblock \emph{arXiv preprint arXiv:1905.08446}, 2019{\natexlab{c}}.

\bibitem[Wang and Samworth(2018)]{wang2018}
Tengyao Wang and Richard~J Samworth.
\newblock High dimensional change point estimation via sparse projection.
\newblock \emph{Journal of the Royal Statistical Society, Series B},
  80:\penalty0 57--83, 2018.

\bibitem[Yao(1984)]{yao1984}
Yi-Ching Yao.
\newblock {Estimation of a noisy discrete-time step function: Bayes and
  empirical Bayes approaches}.
\newblock \emph{The Annals of Statistics}, pages 1434--1447, 1984.

\bibitem[Yao(1988)]{yao1988}
Yi-Ching Yao.
\newblock {Estimating the number of change-points via Schwarz' criterion}.
\newblock \emph{Statistics \& Probability Letters}, 6:\penalty0 181--189, 1988.

\bibitem[Yao(1990)]{yao1990asymptotic}
Yi-Ching Yao.
\newblock On the asymptotic behavior of a class of nonparametric tests for a
  change-point problem.
\newblock \emph{Statistics \& Probability Letters}, 9:\penalty0 173--177, 1990.

\bibitem[Yao and Au(1989)]{yao1989}
Yi-Ching Yao and Siu-Tong Au.
\newblock Least-squares estimation of a step function.
\newblock \emph{Sankhy{\=a}: The Indian Journal of Statistics, Series A}, pages
  370--381, 1989.

\bibitem[Yau and Zhao(2016)]{yau2016}
Chun~Yip Yau and Zifeng Zhao.
\newblock Inference for multiple change points in time series via likelihood
  ratio scan statistics.
\newblock \emph{Journal of the Royal Statistical Society, Series B},
  78:\penalty0 895--916, 2016.

\bibitem[Zhang and Siegmund(2007)]{zhang2007}
Nancy~R Zhang and David~O Siegmund.
\newblock A modified {Bayes} information criterion with applications to the
  analysis of comparative genomic hybridization data.
\newblock \emph{Biometrics}, 63:\penalty0 22--32, 2007.

\bibitem[Zhao et~al.(2019)Zhao, Chen, and Lin]{zhao2019}
Zifeng Zhao, Li~Chen, and Lizhen Lin.
\newblock Change-point detection in dynamic networks via graphon estimation.
\newblock \emph{arXiv preprint arXiv:1908.01823}, 2019.

\bibitem[Zou et~al.(2014)Zou, Yin, Feng, and Wang]{zou2014}
Changliang Zou, Guosheng Yin, Long Feng, and Zhaojun Wang.
\newblock Nonparametric maximum likelihood approach to multiple change-point
  problems.
\newblock \emph{The Annals of Statistics}, 42:\penalty0 970--1002, 2014.

\end{thebibliography}
}

\end{document}